\documentclass[12pt,a4paper,english]{article}
\usepackage[T1]{fontenc}
\usepackage[utf8]{inputenc}
\usepackage{units}
\usepackage{amsmath}
\usepackage{graphicx}
\usepackage{amssymb}
\usepackage{esint}

\makeatletter

\newcommand{\lyxdot}{.}

\usepackage{a4wide}
\usepackage{psfrag}
\makeatletter
\renewcommand{\theequation}{\hbox{\normalsize\arabic{section}.\arabic{equation}}}
\@addtoreset{equation}{section}
\renewcommand{\thefigure}{\hbox{\normalsize\arabic{section}.\arabic{figure}}}
\@addtoreset{figure}{section}
\renewcommand{\thetable}{\hbox{\normalsize\arabic{section}.\arabic{table}}}
\@addtoreset{table}{section}
\makeatother

\usepackage{ifpdf}\ifpdf\usepackage{epstopdf}\usepackage[pdftex,ps2pdf,dvips,colorlinks,urlcolor=blue,citecolor=blue,linkcolor=blue]{hyperref}\else
\usepackage[hypertex,ps2pdf,dvips,colorlinks,urlcolor=blue,citecolor=blue,linkcolor=blue]{hyperref}\fi\pdfadjustspacing=1

\makeatother

\usepackage{babel}

\begin{document}

\title{\begin{flushright}{\normalsize ITP-Budapest Report No. 651}\end{flushright}\vspace{1cm}Sine-Gordon
form factors in finite volume}

\author{G. Fehér\\
 Budapest University of Technology and Economics\\
and\\
 G. Takács\\
 HAS Theoretical Physics Research Group\\
 1117 Budapest, Pázmány Péter sétány 1/A, Hungary}

\date{21th June 2011}
\maketitle
\begin{abstract}
We compare form factors in sine-Gordon theory, obtained via the bootstrap,
to finite volume matrix elements computed using the truncated conformal
space approach. For breather form factors, this is essentially a straightforward
application of a previously developed formalism that describes the
volume dependence of operator matrix elements up to corrections exponentially
decaying with the volume. In the case of solitons, it is necessary
to generalize the formalism to include effects of non-diagonal scattering.
In some cases it is also necessary to take into account some of the
exponential corrections (so-called $\mu$-terms) to get agreement
with the numerical data. For almost all matrix elements the comparison
is a success, with the puzzling exception of some breather matrix
elements that contain disconnected pieces. We also give a short discussion
of the implications of the observed behaviour of $\mu$-terms on the
determination of operator matrix elements from finite volume data,
as occurs e.g. in the context of lattice field theory.
\end{abstract}

\section{Introduction}

The matrix elements of local operators (form factors) are central
objects in quantum field theory. In two-dimensional integrable quantum
field theory the $S$ matrix can be obtained exactly in the framework
of factorized scattering developed in \cite{zam-zam} (for a later
review see \cite{Mussardo:1992uc}). It was shown in \cite{Karowski:1978vz}
that in such theories using the scattering amplitudes as input it
is possible to obtain a set of equations satisfied by the form factors.
The complete system of form factor equations, which provides the basis
for a programmatic approach (the so-called form factor bootstrap)
was proposed in \cite{Kirillov:1987jp}. For a detailed and thorough
exposition of the subject we refer to \cite{Smirnov:1992vz} on bulk
form factors; later this approach was also extended to form factors
of boundary operators \cite{Bajnok:2006ze,Takacs:2008je}. 

Although the connection with the Lagrangian formulation of quantum
field theory is rather indirect in the bootstrap approach, it is thought
that the general solution of the form factor axioms determines the
complete local operator algebra of the theory. This expectation was
confirmed in many cases by explicit comparison of the space of solutions
to the spectrum of local operators as described by the ultraviolet
limiting conformal field theory \cite{Koubek:1993ke,Koubek:1994di,Koubek:1994gk,Koubek:1994zp,Smirnov:1995jp};
the mathematical foundation is provided by the local commutativity
theorem stating that operators specified by solutions of the form
factor bootstrap are mutually local \cite{Smirnov:1992vz}. Another
important piece of information comes from correlation functions: using
form factors, a spectral representation for the correlation functions
can be built which provides a large distance expansion \cite{Yurov:1990kv},
while the Lagrangian or perturbed conformal field theory formulation
allows one to obtain a short-distance expansion, which can then be
compared provided there is an overlap between their regimes of validity
\cite{Zamolodchikov:1990bk}. Other evidence for the correspondence
between the field theory and the solutions of the form factor bootstrap
results from evaluating sum rules like Zamolodchikov's $c$-theorem
\cite{Zamolodchikov:1986gt,Cardy:1988tj,Cappelli:1990yc,Freedman:1990tg}
or the $\Delta$-theorem \cite{Delfino:1996nf}, both of which can
be used to express conformal data as spectral sums in terms of form
factors. Direct comparisons with multi-particle matrix elements are
not so readily available, except for perturbative or $1/N$ calculations
in some simple cases \cite{Karowski:1978vz}. 

Therefore, part of the motivation is to provide non-perturbative evaluation
of form factors from the Hamiltonian formulation, which then allows
for a direct comparison with solutions of the form factor axioms.
Another goal is to have a better understanding of finite size effects
in the case of matrix elements of local operators, and to contribute
to the investigation of finite volume \cite{Smirnov:1998kv,korepin-slavnov,Mussardo:2003ji}
(and also finite temperature \cite{Doyon:2006pv}) form factors and
correlation functions.

This program has been successfully pursued in the case of diagonal
scattering theories (those without particle mass degeneracies), both
in the bulk and with boundary \cite{Pozsgay:2007kn,Pozsgay:2007gx,Pozsgay:2009pv,Kormos:2007qx}.
However, an extension to the non-diagonal case is still missing, and
the present work is a step in that direction. It is important to realize
that non-diagonal theories, whose spectra contain some nontrivial
particle multiplets (typically organized into representations of some
group symmetry), such as the $O(3)$ nonlinear sigma model are very
important for condensed matter applications (e.g. to spin chains;
cf. \cite{Essler:2004ht}). The finite volume description of form
factors can be used to develop a low-temperature and large-distance
expansion for finite-temperature correlation functions \cite{Pozsgay:2007gx,Essler:2009zz},
which could in turn be used to explain experiments such as data from
inelastic neutron scattering \cite{Essler:2007jp}. Therefore the
extension to non-diagonal theories is an interesting direction.

Sine-Gordon model can be considered as the prototype of a non-diagonal
scattering theory, and it has the advantage that its finite volume
spectra and form factors can be studied numerically using the truncated
conformal space approach, originally developed by Yurov and Zamolodchikov
for the scaling Lee-Yang model, but later extended to the sine-Gordon
theory \cite{Feverati:1998va}. Its exact form factors are also known
in full generality (at least in principle -- see the discussion on
multi-soliton form factors later), and so it is a useful playground
to test our theoretical ideas on finite volume form factors. This
cuts in another way too since our approach also provides a way to
test the conjectured exact form factors in much more details than
allowed by the usual methods listed above.

The outline of the paper is as follows. We collect the necessary facts
about the sine-Gordon form factors in section 2. In Section 3 we turn
our attention to the breather form factors, which can be treated by
the methods developed for diagonal scattering theories. Section 4
contains our results on soliton form factors, as well as a conjecture
how the finite volume form factor formulae derived earlier in \cite{Pozsgay:2007kn,Pozsgay:2007gx}
can be extended to non-diagonal theories. At present we are only able
to perform a partial check of this conjecture due to difficulties
in evaluating multi-soliton form factors numerically. Section 5 sums
up our conclusions. The paper also contains an appendix devoted to
a brief description of the algebraic Bethe Ansatz technique that can
be used to obtain the finite volume description of multi-soliton energy
levels.

\section{Brief review of sine-Gordon form factors}

\subsection{Action and $S$ matrix}

The classical action of the theory is \[
\mathcal{A}=\int d^{2}x\left(\frac{1}{2}\partial_{\mu}\Phi\partial^{\mu}\Phi+\frac{m_{0}^{2}}{\beta^{2}}\cos\beta\Phi\right)\]
The fundamental excitations are a doublet of soliton/antisoliton of
mass $M$. Their exact $S$ matrix can be written as \cite{zam-zam}
\begin{equation}
\mathcal{S}_{i_{1}i_{2}}^{j_{1}j_{2}}(\theta,\xi)=S_{i_{1}i_{2}}^{j_{1}j_{2}}(\theta,\xi)S_{0}(\theta,\xi)\label{eq:sg_smatrix}\end{equation}
where\begin{eqnarray*}
 &  & S_{++}^{++}(\theta,\xi)=S_{--}^{--}(\theta,\xi)=1\\
 &  & S_{+-}^{+-}(\theta,\xi)=S_{-+}^{-+}(\theta,\xi)=S_{T}(\theta,\xi)\\
 &  & S_{+-}^{-+}(\theta,\xi)=S_{-+}^{+-}(\theta,\xi)=S_{R}(\theta,\xi)\end{eqnarray*}
and\begin{eqnarray*}
 &  & S_{T}(\theta,\xi)=\frac{\sinh\left(\frac{\theta}{\xi}\right)}{\sinh\left(\frac{i\pi-\theta}{\xi}\right)}\qquad,\qquad S_{R}(\theta,\xi)=\frac{i\sin\left(\frac{\pi}{\xi}\right)}{\sinh\left(\frac{i\pi-\theta}{\xi}\right)}\\
 &  & S_{0}(\theta,\xi)=-\exp\left\{ -i\int_{0}^{\infty}\frac{dt}{t}\frac{\sinh\frac{\pi(1-\xi)t}{2}}{\sinh\frac{\pi\xi t}{2}\cosh\frac{\pi2}{2}}\sin\theta t\right\} \\
 &  & =-\left(\prod_{k=1}^{n}\frac{ik\pi\xi-\theta}{ik\pi\xi+\theta}\right)\exp\Bigg\{-i\int_{0}^{\infty}\frac{dt}{t}\sin\theta t\\
 &  & \quad\times\frac{\left[2\sinh\frac{\pi(1-\xi)t}{2}\mathrm{e}^{-n\pi\xi t}+\left(\mathrm{e}^{-n\pi\xi t}-1\right)\left(\mathrm{e}^{\pi(1-\xi)t/2}+\mathrm{e}^{-\pi(1+\xi)t/2}\right)\right]}{2\sinh\frac{\pi\xi t}{2}\cosh\frac{\pi2}{2}}\Bigg\}\end{eqnarray*}
(the latter representation is valid for any value of $n\in\mathbb{N}$
and makes the integral representation converge faster and further
away from the real $\theta$ axis). 

Besides the solitons, the spectrum of theory contains also breathers
$B_{r}$, with masses\begin{equation}
m_{r}=2M\sin\frac{r\pi\xi}{2}\label{eq:breather_mass}\end{equation}
The breather-soliton and breather-breather $S$-matrices are also
known, here we only quote the $B_{1}-B_{1}$ amplitude that is needed
in the sequel:\begin{equation}
\mathcal{S}_{B_{1}B_{1}}(\theta)=\frac{\sinh\theta+i\sin\pi\xi}{\sinh\theta-i\sin\pi\xi}\label{eq:brbrSmatrix}\end{equation}
Another representation of the theory is as a free massless boson conformal
field theory (CFT) perturbed by a relevant operator. The Hamiltonian
can be written as\begin{equation}
H=\int dx\frac{1}{2}:\left(\partial_{t}\Phi\right)^{2}+\left(\partial_{x}\Phi\right)^{2}:+\mu\int dx:\cos\beta\Phi:\label{eq:pcft_action}\end{equation}
where the semicolon denotes normal ordering in terms of the modes
of the $\mu=0$ massless field. In this case, due to anomalous dimension
of the normal ordered cosine operator, the coupling constant has dimension\[
\mu\sim\left[\mbox{mass}\right]^{2-\beta^{2}/4\pi}\]

\subsection{Breather form factors}

We consider only exponentials of the bosonic field $\Phi$. To obtain
matrix elements containing the first breather, one can analytically
continue the form factors of sinh-Gordon theory obtained in \cite{Koubek:1993ke}
to imaginary values of the couplings. The result is\begin{eqnarray}
F_{\underbrace{{\scriptstyle 11\dots1}}_{n}}^{a}(\theta_{1},\dots,\theta_{n}) & = & \left\langle 0\left|\mathrm{e}^{ia\beta\Phi(0)}\right|B_{1}(\theta_{1})\dots B_{1}(\theta_{n})\right\rangle \nonumber \\
 & = & \mathcal{G}_{a}(\beta)\,[a]_{\xi}\,(i\bar{\lambda}(\xi))^{n}\,\prod_{i<j}\frac{f_{\xi}(\theta_{j}-\theta_{i})}{\mathrm{e}^{\theta_{i}}+\mathrm{e}^{\theta_{j}}}\, Q_{a}^{(n)}\left(\mathrm{e}^{\theta_{1}},\dots,\mathrm{e}^{\theta_{n}}\right)\label{eq:b1ffs}\end{eqnarray}
where \begin{eqnarray*}
Q_{a}^{(n)}(x_{1},\dots,x_{n}) & = & \det{[a+i-j]_{\xi}\,\sigma_{2i-j}^{(n)}(x_{1},\dots,x_{n})}_{i,j=1,\dots,n-1}\mbox{ if }n\geq2\\
Q_{a}^{(1)} & = & 1\quad,\qquad[a]_{\xi}=\frac{\sin\pi\xi a}{\sin\pi\xi}\\
\bar{\lambda}(\xi) & = & 2\cos\frac{\pi\xi}{2}\sqrt{2\sin\frac{\pi\xi}{2}}\exp\left(-\int_{0}^{\pi\xi}\frac{dt}{2\pi}\frac{t}{\sin t}\right)\end{eqnarray*}
and \begin{eqnarray}
f_{\xi}(\theta) & = & v(i\pi+\theta,-1)v(i\pi+\theta,-\xi)v(i\pi+\theta,1+\xi)\nonumber \\
 &  & v(-i\pi-\theta,-1)v(-i\pi-\theta,-\xi)v(-i\pi-\theta,1+\xi)\nonumber \\
v(\theta,\zeta) & = & \prod_{k=1}^{N}\left(\frac{\theta+i\pi(2k+\zeta)}{\theta+i\pi(2k-\text{\ensuremath{\zeta}})}\right)^{k}\exp\Bigg\{\int_{0}^{\infty}\frac{dt}{t}\Big(-\frac{\zeta}{4\sinh\frac{t}{2}}-\frac{i\text{\ensuremath{\zeta}}\theta}{2\pi\cosh\frac{t}{2}}\nonumber \\
 &  & +\left(N+1-N\mbox{e}^{-2t}\right)\mbox{e}^{-2Nt+\frac{it\theta}{\pi}}\frac{\sinh\zeta t}{2\sinh^{2}t}\Big)\Bigg\}\label{eq:brminff}\end{eqnarray}
gives the minimal $B_{1}B_{1}$ form factor%
\footnote{The formula for the function $v$ is in fact independent of $N$;
choosing $N$ large extends the width of the strip where the integral
converges and also speeds up convergence.%
}, while $\sigma_{k}^{(n)}$ denotes the elementary symmetric polynomial
of $n$ variables and order $k$ defined by\[
\prod_{i=1}^{n}(x+x_{i})=\sum_{k=0}^{n}x^{n-k}\sigma_{k}^{(n)}(x_{1},\dots,x_{n})\]
Furthermore\begin{eqnarray}
\mathcal{G}_{a}(\beta)=\langle e^{ia\beta\Phi}\rangle & = & \left[\frac{M\sqrt{\pi}\Gamma\left(\frac{4\pi}{8\pi-\beta^{2}}\right)}{2\Gamma\left(\frac{\beta^{2}/2}{8\pi-\beta^{2}}\right)}\right]^{\frac{a^{2}\beta^{2}}{4\pi}}\exp\Bigg\{\int_{0}^{\infty}\frac{dt}{t}\Bigg[-\frac{a^{2}\beta^{2}}{4\pi}e^{-2t}\nonumber \\
 &  & +\frac{\sinh^{2}\left(\frac{a}{4\pi}t\right)}{2\sinh\left(\frac{\beta^{2}}{8\pi}t\right)\cosh\left(\left(1-\frac{\beta^{2}}{8\pi}\right)t\right)\sinh t}\Bigg]\Bigg\}\label{eq:exactvev}\end{eqnarray}
is the exact vacuum expectation value of the exponential field \cite{Lukyanov:1996jj},
with $M$ denoting the soliton mass related to the coupling $\mu$
defined in via \cite{Zamolodchikov:1995xk} \begin{equation}
\mu=\frac{2\Gamma(\Delta)}{\pi\Gamma(1-\Delta)}\left(\frac{\sqrt{\pi}\Gamma\left(\frac{1}{2-2\Delta}\right)M}{2\Gamma\left(\frac{\Delta}{2-2\Delta}\right)}\right)^{2-2\Delta}\qquad,\qquad\Delta=\frac{\beta^{2}}{8\pi}\label{eq:mass_scale}\end{equation}
Formula (\ref{eq:b1ffs}) also coincides with the result given in
\cite{Lukyanov:1997bp}. Form factors of higher breathers can be constructed
by considering them as bound states of $B_{1}$ particles. Detailed
formulae can be found in Appendix A of \cite{Takacs:2009fu}.

\subsection{Soliton form factors}

At present, there are three independent constructions of solitonic
form factors available: the earliest one by Smirnov (reviewed in \cite{smirnov_ff}),
the free field representation by Lukyanov \cite{Lukyanov:1993pn,Lukyanov:1997bp}
and the work by Babujian et al. \cite{Babujian:1998uw,Babujian:2001xn}.
Here we use formulae from Lukyanov's work \cite{Lukyanov:1997bp}.
The simplest form factor for an exponential operator is with a soliton-antisoliton
state and it takes the form\begin{align}
\left\langle 0\left|\mathrm{e}^{i\beta\Phi(0)}\right|S_{\pm}(\theta_{2})S_{\mp}(\theta_{1})\right\rangle  & =F_{\mp\pm}^{\beta}(\theta_{1}-\theta_{2})\nonumber \\
F_{\mp\pm}^{\beta}(\theta) & =\mathcal{G}_{1}(\beta)\, G(\theta)\cot\left(\frac{\pi\xi}{2}\right)\frac{4i\cosh\left(\frac{\theta}{2}\right)\mathrm{e}^{\mp\frac{\theta+i\pi}{2\xi}}}{\xi\sinh\left(\frac{\theta+i\pi}{\xi}\right)}\label{eq:twossff}\end{align}
where $S_{+}$ and $S_{-}$ denote the soliton and antisoliton, respectively.
The function $G$ is given by the integral representation\begin{eqnarray*}
G(\theta) & = & i\sinh\left(\frac{\theta}{2}\right)\exp\left\{ \int_{0}^{\infty}\frac{dt}{t}\sinh^{2}\left(\left(1-\frac{i\theta}{\pi}\right)t\right)\frac{\sinh\left(t\left(\xi-1\right)\right)}{\sinh\left(2t\right)\cosh\left(t\right)\sinh\left(t\xi\right)}\right\} \\
 & = & i\sinh\left(\frac{\theta}{2}\right)\prod_{k=1}^{N}g(\theta,\xi,k)^{k}\,\exp\Bigg\{\int_{0}^{\infty}\frac{dt}{t}\mathrm{e}^{-4Nt}\left(1+N-N\,\mathrm{e^{-4t}}\right)\\
 &  & \times\sinh^{2}\left(\left(1-\frac{i\theta}{\pi}\right)t\right)\frac{\sinh\left(t\left(\xi-1\right)\right)}{\sinh\left(2t\right)\cosh\left(t\right)\sinh\left(t\xi\right)}\Bigg\}\\
g(\theta,\xi,k) & = & \frac{\Gamma\left(\frac{(2k+1+\xi)\pi-i\theta}{\pi\xi}\right)\Gamma\left(\frac{2k+1}{\xi}\right)^{2}\Gamma\left(\frac{(2k+1)\pi-i\theta}{\pi\xi}\right)}{\Gamma\left(\frac{2k+\xi}{\xi}\right)^{2}\Gamma\left(\frac{(2k+\xi)\pi-i\theta}{\pi\xi}\right)\Gamma\left(\frac{(2k-2+\xi)\pi+i\theta}{\pi\xi}\right)}\\
 &  & \times\frac{\Gamma\left(\frac{(2k-1)\pi+i\theta}{\pi\xi}\right)\Gamma\left(\frac{2k-1+\xi}{\xi}\right)^{2}\Gamma\left(\frac{(2k-1+\xi)\pi+i\theta}{\pi\xi}\right)}{\Gamma\left(\frac{2k}{\xi}\right)^{2}\Gamma\left(\frac{(2k+2)\pi-i\theta}{\pi\xi}\right)\Gamma\left(\frac{2k\pi+i\theta}{\pi\xi}\right)}\end{eqnarray*}
where, again, the second formula is eventually independent of the
natural number $N$; it provides a representation which converges
faster numerically and is valid further away from the real $\theta$
axis with increasing $N$.

Higher form factors in all the available constructions are given in
some quite complicated integral representation which we do not write
down here. We have been unable to find a numerical evaluation of these
analytic formulae at present, which precludes their use in a comparison
with TCSA data.

\section{Breather form factors in finite volume}

\subsection{A review of the theoretical predictions}

As breathers are singlet and so scatter diagonally, the formulae presented
in the papers \cite{Pozsgay:2007kn,Pozsgay:2007gx} are directly applicable.
The first ingredient is to describe the multi-breather energy levels
corresponding to the states\[
\left|B_{r_{1}}(\theta_{1})\dots B_{r_{N}}(\theta_{N})\right\rangle \]
whose finite volume counter part we are going to label\[
\vert\{I_{1},\dots,I_{N}\}\rangle_{r_{1}\dots r_{N},L}\]
where the $I_{k}$ are the momentum quantum numbers. In a finite volume
$L$, momentum quantization is governed (up to corrections decreasing
exponentially with $L$) by the Bethe-Yang equations:\begin{equation}
Q_{k}(\theta_{1},\dots,\theta_{n})=m_{r_{k}}L\sinh\theta_{k}+\sum_{j\neq k}\delta_{r_{j}r_{k}}\left(\theta_{k}-\theta_{j}\right)=2\pi I_{k}\qquad I_{k}\in\mathbb{Z}\label{eq:breatherby}\end{equation}
where the phase-shift is defined as \[
S_{B_{r}B_{s}}(\theta)=\mathrm{e}^{i\delta_{sr}(\theta)}\]
In practical calculations, because breathers of the same species satisfy
an effective exclusion rule due to\[
S_{B_{r}B_{r}}(0)=-1\]
it is best to redefine phase-shifts corresponding to them by extracting
a $-$ sign:\[
S_{B_{r}B_{r}}(\theta)=-\mathrm{e}^{i\delta_{rr}(\theta)}\]
so that $\delta_{sr}(0)=0$ can be taken for all $s,r$ and all the
phase-shifts can be defined as continuous functions over the whole
real $\theta$ axis. This entails shifting appropriate quantum numbers
$I_{k}$ to half-integer values. Given a solution $\tilde{\theta}_{1},\dots,\tilde{\theta}_{N}$
to the quantization relations (\ref{eq:breatherby}) the energy and
the momentum of the state can be written as \begin{eqnarray*}
E & = & \sum_{k=1}^{N}m_{r_{k}}\cosh\tilde{\theta}_{k}\\
P & = & \sum_{k=1}^{N}m_{r_{k}}\sinh\tilde{\theta}_{k}=\frac{2\pi}{L}\sum_{k}I_{k}\end{eqnarray*}
(using that -- with our choice of the phase-shift functions -- unitarity
entails $\delta_{sr}(\theta)+\delta_{rs}(-\theta)=0$). The rapidity-space
density of $n$-particle states can be calculated as \begin{equation}
\rho_{r_{1}\dots r_{n}}(\theta_{1},\dots,\theta_{n})=\det\mathcal{J}^{(n)}\qquad,\qquad\mathcal{J}_{kl}^{(n)}=\frac{\partial Q_{k}(\theta_{1},\dots,\theta_{n})}{\partial\theta_{l}}\quad,\quad k,l=1,\dots,n\label{eq:byjacobian}\end{equation}
The matrix elements of local operators between finite volume multi-particle
states can be written as \cite{Pozsgay:2007kn} \begin{eqnarray}
 &  & \left|\,_{s_{1}\dots s_{M}}\langle\{I_{1}',\dots,I_{M}'\}\vert\mathcal{O}(0,0)\vert\{I_{1},\dots,I_{N}\}\rangle_{r_{1}\dots r_{N},L}\right|=\nonumber \\
 &  & \qquad\left|\frac{F_{s_{M}\dots s_{1}r_{1}\dots r_{N}}^{\mathcal{O}}(\tilde{\theta}_{M}'+i\pi,\dots,\tilde{\theta}_{1}'+i\pi,\tilde{\theta}_{1},\dots,\tilde{\theta}_{N})}{\sqrt{\rho_{r_{1}\dots r_{N}}(\tilde{\theta}_{1},\dots,\tilde{\theta}_{N})\rho_{s_{1}\dots s_{M}}(\tilde{\theta}_{1}',\dots,\tilde{\theta}_{M}')}}\right|+O(\mathrm{e}^{-\mu'L})\label{eq:genffrelation}\end{eqnarray}
(note that we cannot specify the phase of the matrix elements as it
depends on phase conventions which can be different in finite and
infinite volume) which is valid provided there are no disconnected
terms. Such terms appear when there are two breathers of the same
species in the two states whose rapidities coincide. Apart from some
very special cases, this happens only when the two states are identical,
which is called the diagonal matrix element. The necessary formulae
are written down in \cite{Pozsgay:2007gx}; instead of quoting here
the rather lengthy general expression, we present the two particular
cases we need:\begin{eqnarray}
\,_{1}\langle\{I\}\vert\mathcal{O}(0,0)\vert\{I\}\rangle_{1,L} & = & \frac{F_{11}^{\mathcal{O}}(i\pi,0)}{m_{1}L\cosh\tilde{\theta}}+\left\langle \mathcal{O}\right\rangle +O(\mathrm{e}^{-\mu'L})\label{eq:br_diagff}\\
\,_{11}\langle\{I_{1},I_{2}\}\vert\mathcal{O}(0,0)\vert\{I_{1},I_{2}\}\rangle_{11,L} & = & \frac{\mathcal{F}_{11}^{\mathcal{O}}(\tilde{\theta}_{1},\tilde{\theta}_{2})+m_{1}L\left(\cosh\tilde{\theta}_{1}+\cosh\tilde{\theta}_{2}\right)F_{11}^{\mathcal{O}}(i\pi,0)}{\rho_{11}(\tilde{\theta}_{1},\tilde{\theta}_{2})}\nonumber \\
 &  & +\left\langle \mathcal{O}\right\rangle +O(\mathrm{e}^{-\mu'L})\nonumber \end{eqnarray}
where\[
\left\langle \mathcal{O}\right\rangle \]
is the vacuum expectation value of the local operator $\mathcal{O}$
and \[
\mathcal{F}_{11}^{\mathcal{O}}(\theta_{1},\theta_{2})=\lim_{\epsilon\rightarrow0}F_{1111}^{\mathcal{O}}(\theta_{2}+i\pi+\epsilon,\theta_{1}+i\pi+\epsilon,\theta_{1},\theta_{2})\]
is the so-called symmetric evaluation of the four-particle form factor.
Note that in the diagonal case there is no possible phase difference
between the finite-volume matrix elements and the infinite volume
form factors, as any phase redefinition of the state drops out from
the matrix element.

The above predictions for finite volume energy levels and matrix elements
are expected to be exact to all (finite) orders in $1/L$ \cite{Pozsgay:2007kn,Pozsgay:2007gx}
(note that the exponential corrections are non-analytic in this variable).

\subsection{Numerical results}

To evaluate the form factors numerically, we use the truncated conformal
space approach (TCSA) pioneered by Yurov and Zamolodchikov \cite{Yurov:1989yu}.
The extension to the sine-Gordon model was developed in \cite{Feverati:1998va}
and has found numerous applications since then. The Hilbert space
can be split by the eigenvalues of the topological charge $\mathcal{Q}$
(or winding number) and the spatial momentum $P$, where the eigenvalues
of the latter are of the form\[
\frac{2\pi s}{L}\]
$s$ is called the 'conformal spin'. 

In sectors with vanishing topological charge, we can make use of the
symmetry of the Hamiltonian under \[
\mathcal{C}:\qquad\Phi(x,t)\rightarrow-\Phi(x,t)\]
which is equivalent to conjugation of the solitonic charge. The truncated
space can be split into $\mathcal{C}$-even and $\mathcal{C}$-odd
subspaces that have roughly equal dimensions, which speeds up the
diagonalization of the Hamiltonian by roughly a factor of eight (the
required machine time scales approximately with the third power of
matrix size). We remark that there is another discrete symmetry\[
\mathcal{P}:\qquad\Phi(x,t)\rightarrow-\Phi(-x,t)\]
corresponding to spatial parity, which is a symmetry in $s=0$ sectors.
However, we use the $s\neq0$ states extensively in our calculations. 

For the calculations, we choose the operator\[
\mathcal{O}=:\mathrm{e}^{i\beta\Phi}:\]
which is essentially one half of the interaction term in the Hamiltonian
in (\ref{eq:pcft_action}). The semicolons denote normal ordering
with respect to the $\lambda=0$ free massless boson modes. This operator
has conformal dimension\[
\Delta_{\mathcal{O}}=\bar{\Delta}_{\mathcal{O}}=\frac{\beta^{2}}{8\pi}\]
Using relation (\ref{eq:mass_scale}) we can express all energy levels
and matrix elements in units of (appropriate powers of) the soliton
mass $M$, and we also introduce the dimensionless volume variable
$l=ML$. The general procedure is the same as in \cite{Pozsgay:2007kn,Pozsgay:2007gx}:
the particle content of energy levels can be identified by matching
the numerical TCSA spectrum against the predictions of the Bethe-Yang
equations (\ref{eq:breatherby}). After identification, one can compare
the appropriate matrix elements to the theoretical values given by
(\ref{eq:genffrelation}) and (\ref{eq:br_diagff}).

Due to level crossings, at certain values of the volume $L$ there
can be more than one TCSA candidate levels for a given Bethe-Yang
solution; identification can be completed by selecting the candidate
on the basis of one of the form factor measurements, which still leaves
other matrix elements involving the state as cross-checks. Level crossings
also present a problem in numerical stability, since in their vicinity
the state of interest is nearly degenerate to another one. Since the
truncation effect can be considered as an additional perturbing operator,
the level crossings are eventually lifted. However, such a near-degeneracy
greatly magnifies truncation effects on the eigenvectors and therefore
the matrix elements \cite{Kormos:2007qx}. This is the reason behind
the fact that in some of the figures there are some individual numerical
points that are clearly scattered away from their expected place.

The simplest matrix element to test is the theoretical prediction
(\ref{eq:exactvev}) for the exact vacuum expectation value, and we
did check it against the numerical results. However, it has already
been subjected to extensive verification against TCSA in \cite{Bajnok:2000wm},
therefore we do not dwell on this issue here. 

\begin{figure}
\begin{centering}
\psfrag{Volume}{$l$}\psfrag{formfactor}{$|F_n|$}
\psfrag{ff1TCSA}{n=1 TCSA}\psfrag{ff2TCSA}{n=2 TCSA}\psfrag{ff3TCSA}{n=3 TCSA}
\psfrag{ff1THEORY}{n=1 theory}\psfrag{ff2THEORY}{n=2 theory}\psfrag{ff3THEORY}{n=3 theory}\includegraphics[scale=0.7]{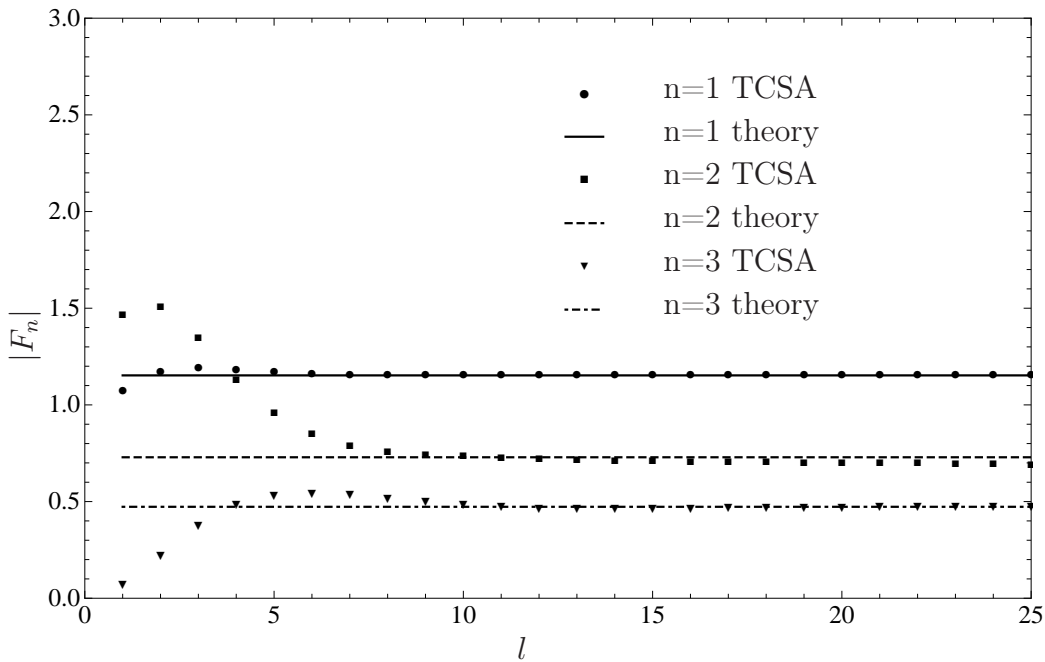}
\par\end{centering}

\caption{\label{fig:One-B123-form-factors} One-particle form factors at $\xi=50/239$}
\end{figure}

One-particle form factors can be tested using a reorganized form of
the relation (\ref{eq:genffrelation}):\[
\left|F_{n}^{\mathcal{O}}\right|=\left|\left\langle 0\left|\mathrm{e}^{i\beta\Phi(0)}\right|B_{n}(0)\right\rangle \right|=\left|\sqrt{m_{n}L}\left\langle 0\left|\mathrm{e}^{i\beta\Phi(0)}\right|\{0\}\right\rangle _{n,L}\right|+O(\mathrm{e}^{-\mu'L})\]
(the form factors themselves are independent of the rapidity of the
particle due to Lorentz invariance). For the lowest three breathers
the agreement between numerics and theory is illustrated in figure
\ref{fig:One-B123-form-factors}.

\begin{figure}
\begin{centering}
\psfrag{Rapidity}{$\theta$}\psfrag{formfactor}{$|F_{11}(\theta)|$}
\psfrag{ffb1b1RapTHEORY}{theory}
\psfrag{ffs12m12TCSA}{from $\langle 0|\mathcal{O}|\{-1/2,1/2\}\rangle_{11,L}$}
\psfrag{ffs32m32TCSA}{from $\langle 0|\mathcal{O}|\{-3/2,3/2\}\rangle_{11,L}$}
\psfrag{ffs52m52TCSA}{from $\langle 0|\mathcal{O}|\{-5/2,5/2\}\rangle_{11,L}$}\includegraphics[scale=0.8]{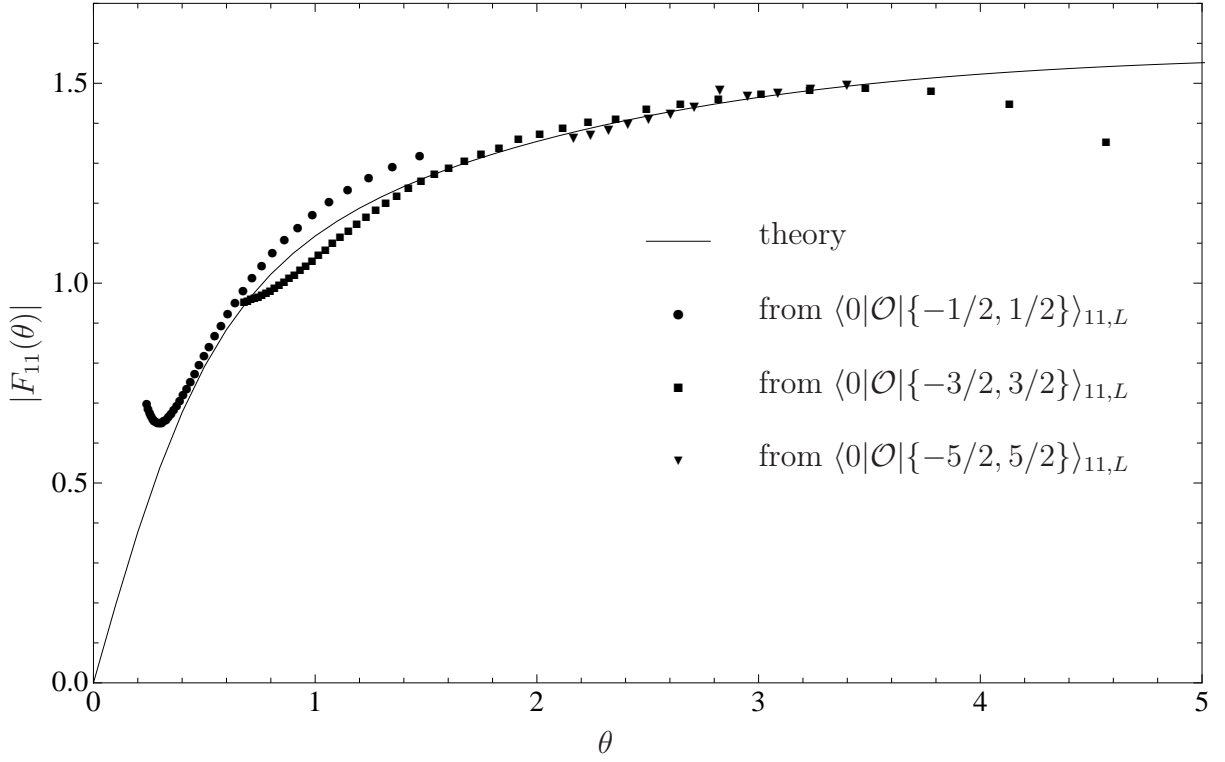}
\par\end{centering}

\begin{centering}
\psfrag{Rapidity}{$\theta$}\psfrag{formfactor}{$|F_{11}(\theta)|$}
\psfrag{ffb1b1RapTHEORY}{theory}
\psfrag{ffs32m12TCSA}{from $\langle 0|\mathcal{O}|\{-1/2,3/2\}\rangle_{11,L}$}
\psfrag{ffs52m32TCSA}{from $\langle 0|\mathcal{O}|\{-3/2,5/2\}\rangle_{11,L}$}\includegraphics[scale=0.8]{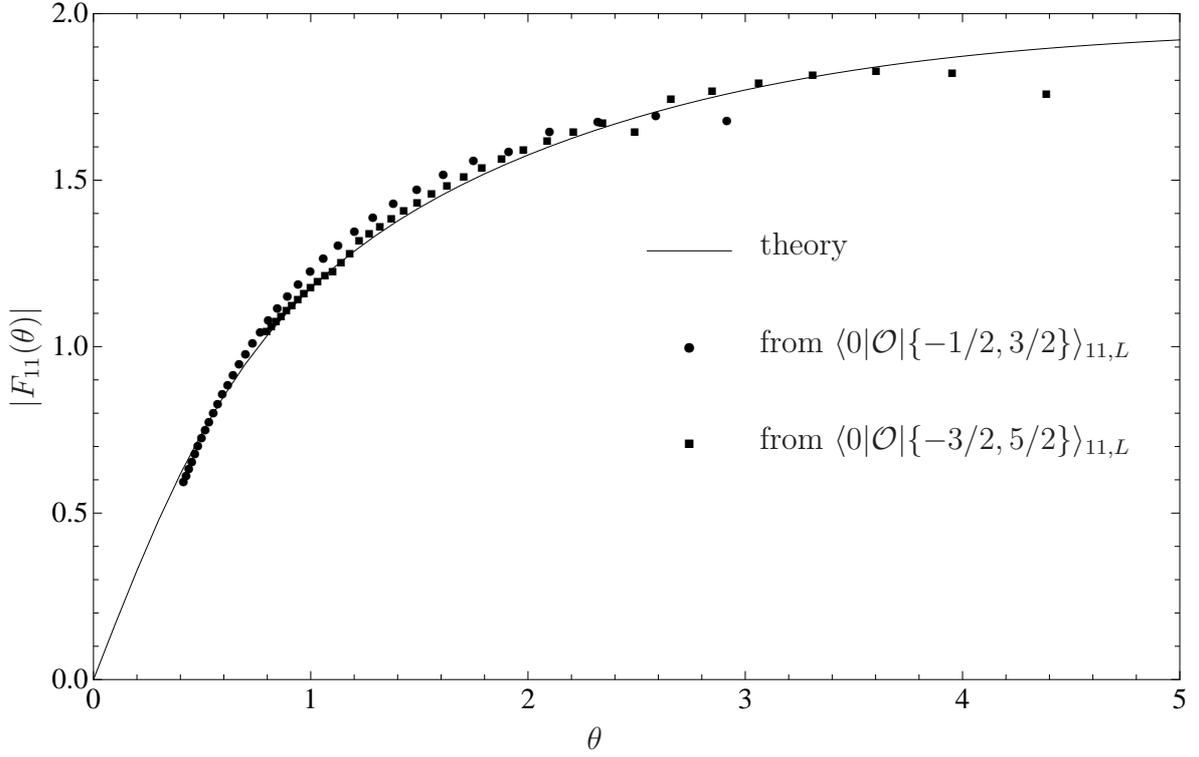}
\par\end{centering}

\caption{\label{fig:b1b1-two-particle-form} $B_{1}B_{1}$ two-particle form
factors. The first plot shows spin-$0$ data with $\xi=50/239$, while
the second one shows spin-1 data with $\xi=2/7$.}

\end{figure}

\begin{figure}

\begin{centering}
\psfrag{Rapidity}{$\theta$}\psfrag{formfactor}{$|F_{11}(i\pi+\theta)|$}
\psfrag{ffb1s0Eb1s1RapTHEORY}{theory}
\psfrag{ffb1s0Eb1s1TCSA}{from TCSA}\includegraphics[scale=0.7]{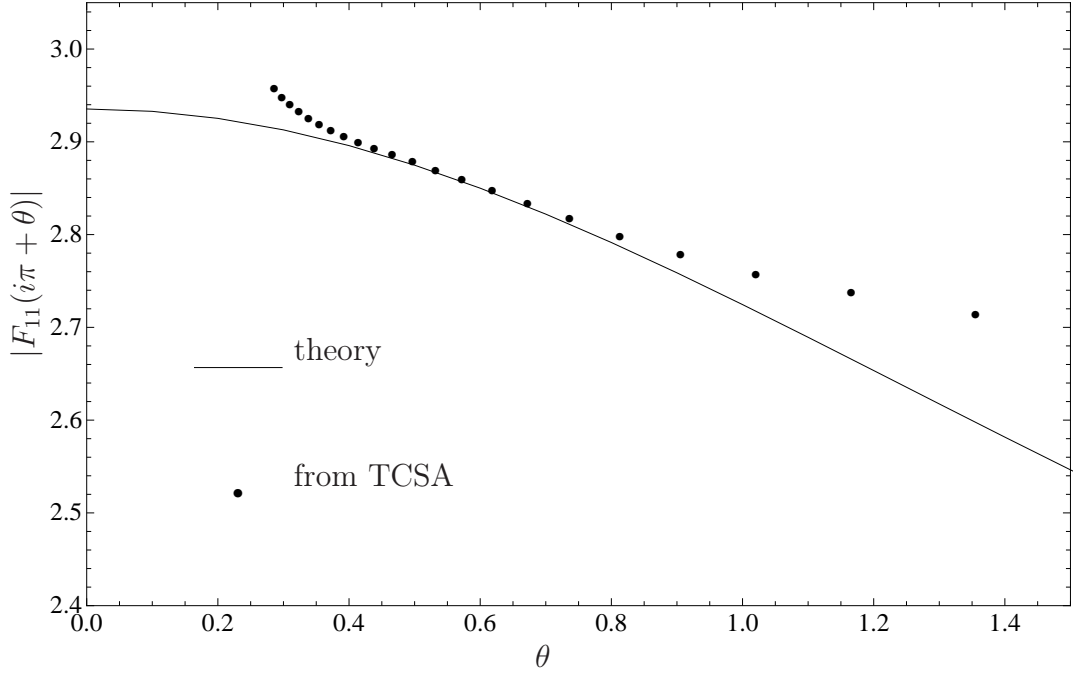}
\par\end{centering}

\caption{\label{fig:Crossedb1b1} Crossed $B_{1}B_{1}$ two-particle form factor.
The TCSA measurements are taken from the matrix element $\,_{1}\langle\{1\}|\mathcal{O}|\{0\}\rangle_{1}$
at $\xi=2/7$.}
\end{figure}

\begin{figure}
\begin{centering}
\psfrag{Volume}{$l$}\psfrag{formfactor}{${}_1\langle{I}|\mathcal{O}|{I}\rangle_{1,L}$}
\psfrag{ffb1s0Eb1s0DisconTCSA}{$I=0$, TCSA}
\psfrag{ffb1s0Eb1s0DisconTHEORY}{$I=0$, theory}
\psfrag{ffb1s1Eb1s1DisconTCSA}{$I=1$, TCSA}
\psfrag{ffb1s1Eb1s1DisconTHEORY}{$I=1$, theory}\includegraphics[scale=0.7]{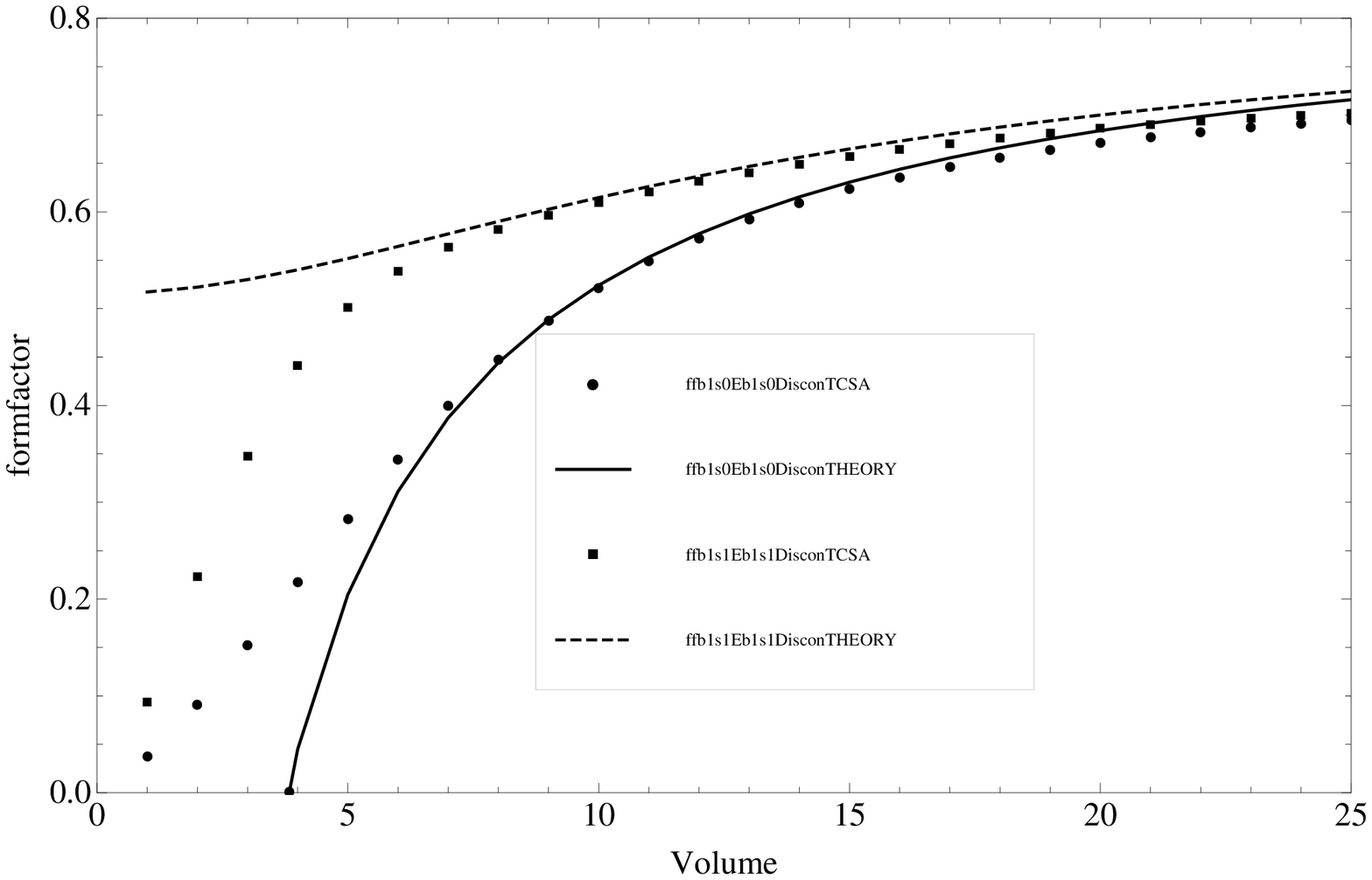}
\par\end{centering}

\caption{\label{fig:Diagonalb1b1} Diagonal $B_{1}$-$B_{1}$ matrix elements
at $\xi=50/239$.}

\end{figure}

The next interesting matrix element is the two-particle one, for which
we can use the relation\[
\left|F_{11}^{\mathcal{O}}(\tilde{\theta}_{1},\tilde{\theta}_{2})\right|=\left|\left\langle 0\left|\mathrm{e}^{i\beta\Phi(0)}\right|B_{1}(\tilde{\theta}_{1})B_{1}(\tilde{\theta}_{2})\right\rangle \right|=\left|\sqrt{\rho_{11}(\tilde{\theta}_{1},\tilde{\theta}_{2})}\left\langle 0\left|\mathrm{e}^{i\beta\Phi(0)}\right|\{I_{1},I_{2}\}\right\rangle _{11,L}\right|+O(\mathrm{e}^{-\mu'L})\]
which is tested in figure \ref{fig:b1b1-two-particle-form}. This
provides a test of the two-particle form factor for real rapidity
differences. However, the two-particle form factor function also appears
in the (off-diagonal and diagonal) $B_{1}$--$B_{1}$ matrix elements\begin{eqnarray*}
\left|_{1}\left\langle \{I'\}\left|\mathrm{e}^{i\beta\Phi(0)}\right|\{I\}\right\rangle _{1,L}\right| & = & \frac{\left|F_{11}^{\mathcal{O}}(i\pi+\tilde{\theta}',\tilde{\theta})\right|}{m_{1}L\sqrt{\cosh\tilde{\theta}'\cosh\tilde{\theta}}}+O(\mathrm{e}^{-\mu'L})\\
_{1}\left\langle \{I\}\left|\mathrm{e}^{i\beta\Phi(0)}\right|\{I\}\right\rangle _{1,L} & = & \frac{F_{11}^{\mathcal{O}}(i\pi,0)}{m_{1}L\cosh\tilde{\theta}}+\mathcal{G}_{1}(\beta)+O(\mathrm{e}^{-\mu'L})\end{eqnarray*}
which makes it possible to test the form factor for rapidity differences
with imaginary part $\pi$ as shown in figures \ref{fig:Crossedb1b1}
and \ref{fig:Diagonalb1b1}. We remark that in plots against rapidity
the small-rapidity deviations are due to truncation errors, while
the ones for large rapidity result from the neglected exponential
corrections.

\begin{figure}
\begin{centering}
\psfrag{Volume}{$l$}\psfrag{formfactor}{$|{}_1\langle\{0\}|\mathcal{O}|\{I_1,I_2\}\rangle_{11,L}|$}
\psfrag{b1Eb1b1s12TCSA}{$I_1=-I_2=1/2$, TCSA}
\psfrag{b1Eb1b1s12THEORY}{$I_1=-I_2=1/2$, theory}
\psfrag{b1Eb1b1s32TCSA}{$I_1=-I_2=3/2$, TCSA}
\psfrag{b1Eb1b1s32THEORY}{$I_1=-I_2=3/2$, theory}\includegraphics[scale=0.7]{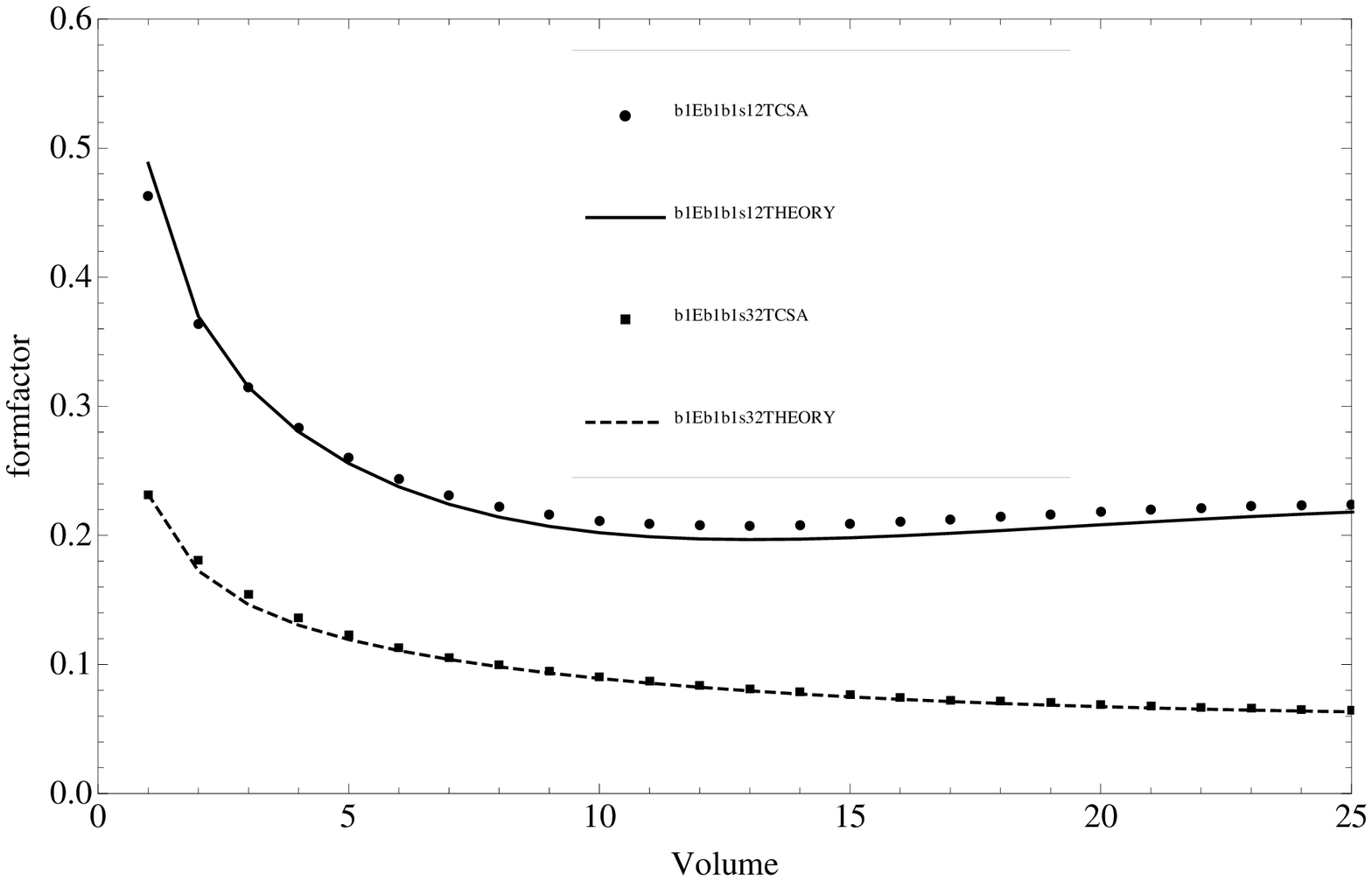}
\par\end{centering}

\caption{\label{fig:b1-b1b1} $B_{1}$-$B_{1}B_{1}$ matrix elements at $\xi=50/239$.}
\end{figure}

\begin{figure}
\begin{centering}
\psfrag{Volume}{$l$}\psfrag{formfactor}{$|{}_{11}\langle\{I',-I'\}|\mathcal{O}|\{I,-I\}\rangle_{11,L}|$}
\psfrag{b1b1s12Eb1b1s32TCSA}{$I=1/2$,$I'=3/2$,TCSA}
\psfrag{b1b1s12Eb1b1s32THEORY}{$I=1/2$,$I'=3/2$,theory}
\psfrag{b1b1s12Eb1b1s52TCSA}{$I=1/2$,$I'=5/2$,TCSA}
\psfrag{b1b1s12Eb1b1s52THEORY}{$I=1/2$,$I'=5/2$,,theory}
\psfrag{b1b1s32Eb1b1s52TCSA}{$I=3/2$,$I'=5/2$,TCSA}
\psfrag{b1b1s32Eb1b1s52THEORY}{$I=3/2$,$I'=5/2$,theory}\includegraphics[scale=0.7]{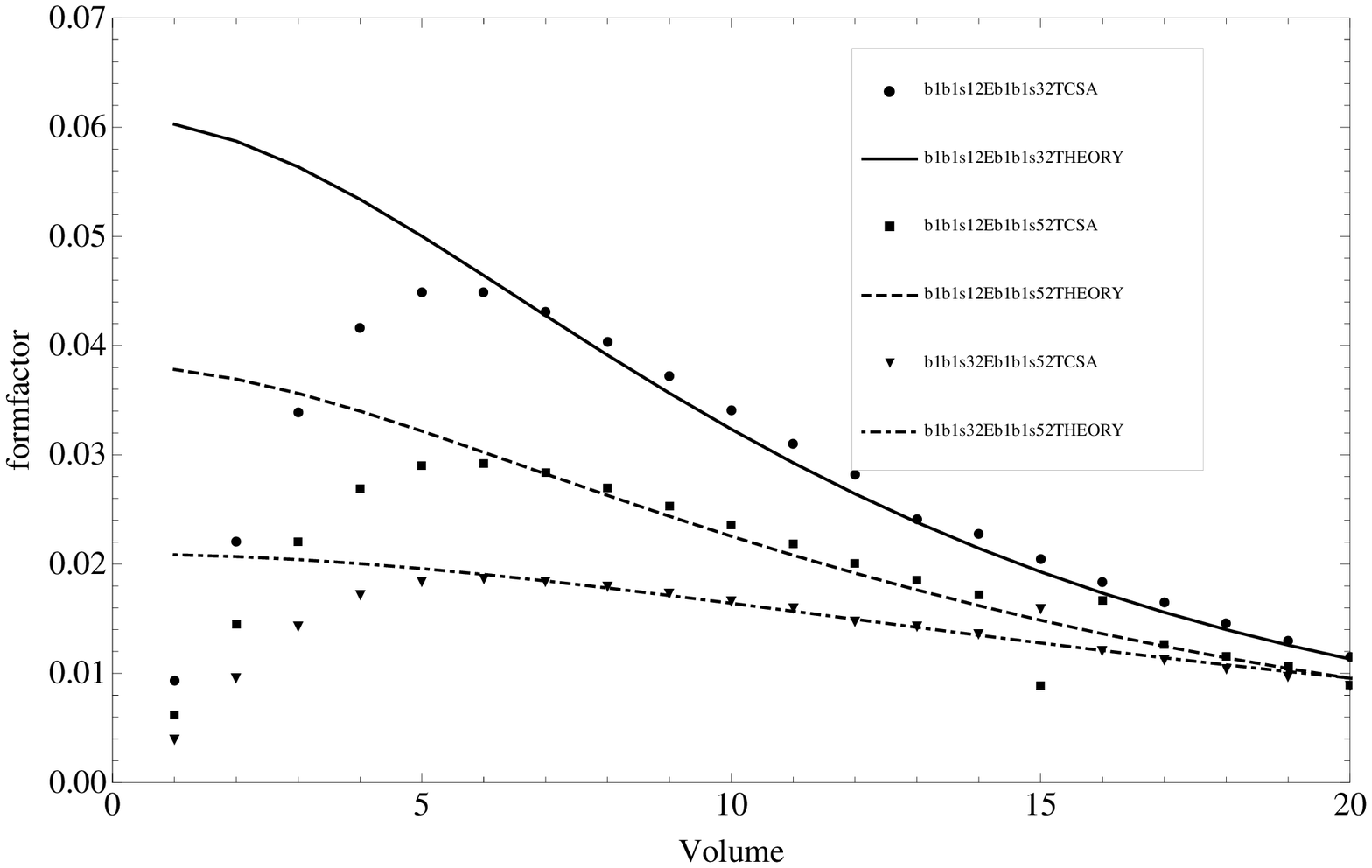}
\par\end{centering}

\caption{\label{fig:b1b1-b1b1offdiag} Off-diagonal $B_{1}B_{1}$-$B_{1}B_{1}$
matrix elements at $\xi=50/239$. The individual deviations are caused
by the proximity of level crossings.}
\end{figure}

We can also test the three and four particle form-factors using $B_{1}-B_{1}B_{1}$
and off-diagonal $B_{1}B_{1}-B_{1}B_{1}$ matrix elements, as shown
in figures \ref{fig:b1-b1b1} and \ref{fig:b1b1-b1b1offdiag}. Diagonal
$B_{1}B_{1}-B_{1}B_{1}$ matrix elements reveal some complications,
and are treated in subsection \ref{sub:Diagonal--matrix}. %

All of these tests show excellent agreement between numerics and theory,
which make us confident both in the finite volume form factor formalism
and the correctness of the exact form factors (\ref{eq:b1ffs}) predicted
by the bootstrap.

\subsection{$B_{2}$ matrix elements and $\mu$-terms \label{sub:mu-terms-for-b2}}

We can also use the identified $B_{2}$ state to compute matrix elements
involving it. However, in some cases simply using (\ref{eq:genffrelation})
gives a rather poor match. It turns out that accounting for some of
the hitherto neglected exponential corrections can improve this; the
leading corrections come from so-called $\mu$-terms which were constructed
in \cite{Pozsgay:2008bf} building upon the ideas in Lüscher's seminal
paper \cite{Luscher:1985dn}. We remark that the method outlined below
eventually does more; it entails a resummation of these correction
beyond the mere construction of the leading exponential correction.

Using the ideas in \cite{Pozsgay:2008bf}, we can model a finite volume
$B_{2}$ state as a pair of $B_{1}$ particles with complex conjugate
rapidities by solving the $B_{1}B_{1}$ Bethe-Yang equations (here
written in exponential form):\begin{align}
\mathrm{e}^{im_{1}L\sinh(\theta\pm iu)}S_{B_{1}B_{1}}(\pm2iu) & =1\label{eq:b2_b1b1_by}\end{align}
The solution for $u$ has the large volume behaviour \begin{equation}
u\sim\frac{\pi\xi}{2}+C\mathrm{e}^{-\mu L\cosh\theta}\label{eq:u_asymptotics}\end{equation}
with \[
\mu=\sqrt{m_{1}^{2}-\frac{m_{2}^{2}}{4}}=m_{1}\sin\frac{\pi\xi}{2}\]
and $C$ some numerical constant. Therefore, for $L\rightarrow\infty$
one obtains the usual bootstrap identification\[
B_{2}(\theta)\simeq B_{1}(\theta+i\pi\xi/2)B_{1}(\theta-i\pi\xi/2)\]
and the momentum and energy of the state tends to\begin{align*}
m_{1}\sinh(\theta+iu)+m_{1}\sinh(\theta-iu) & =2m_{1}\cos u\sinh\theta\mathop{\rightarrow}_{L\rightarrow\infty}m_{2}\sinh\theta\\
m_{1}\cosh(\theta+iu)+m_{1}\cosh(\theta-iu) & =2m_{1}\cos u\cosh\theta\mathop{\rightarrow}_{L\rightarrow\infty}m_{2}\cosh\theta\end{align*}
where \[
m_{2}=2m_{1}\cos\frac{\pi\xi}{2}\]
is the mass of \textbf{$B_{2}$} in terms of that of $B_{1}$, consistent
with (\ref{eq:breather_mass}).

\begin{figure}
\begin{centering}
\psfrag{Volume}{$l$}
\psfrag{formfactor}{$|{}_{2}\langle\{0\}|\mathcal{O}|\Psi\rangle_L|$}
\psfrag{ffb2TCSA}{$|\Psi\rangle=|0\rangle$,TCSA}
\psfrag{ffb2NaiveTHEORY}{$|\Psi\rangle_L=|0\rangle_L$, naive}
\psfrag{ffb2ImRapTheory}{$|\Psi\rangle_L=|0\rangle_L$, $\mu$-terms}
\psfrag{ffb1s0Eb2s0TCSA}{$|\Psi\rangle_L=|\{0\}\rangle_{1,L}$,TCSA}
\psfrag{ffb1s0Eb2s0NaiveTHEORY}{$|\Psi\rangle_L=|\{0\}\rangle_{1,L}$, naive}
\psfrag{ffb1s0Eb2s0ImRapTheory}{$|\Psi\rangle_L=|\{0\}\rangle_{1,L}$, $\mu$-terms}\includegraphics[scale=0.7]{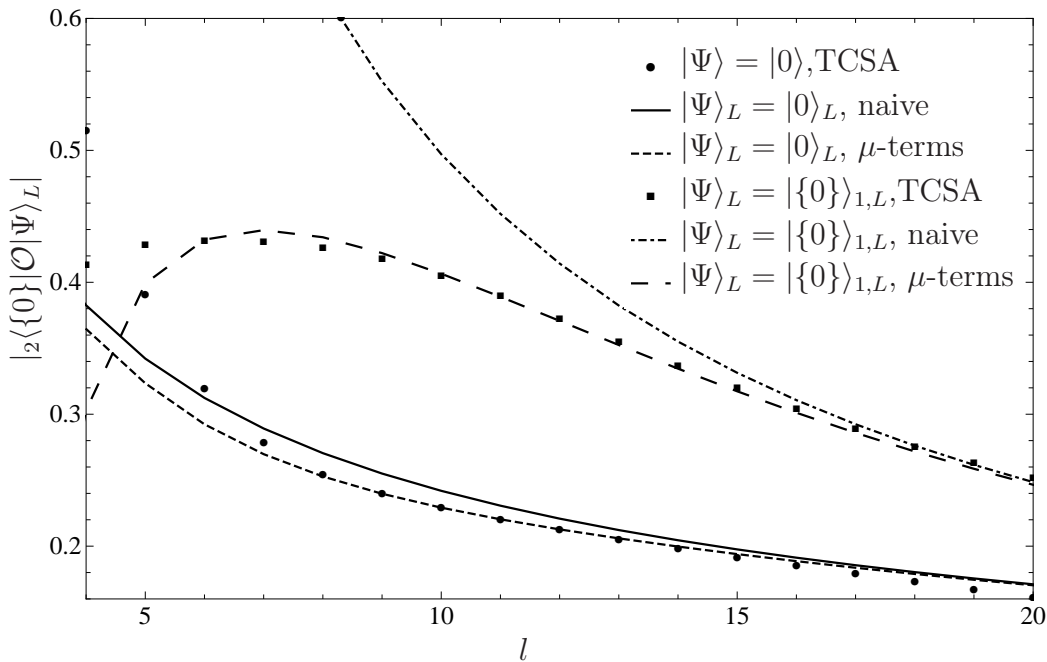}
\par\end{centering}

\caption{\label{fig:mu-terms} Naive predictions for matrix elements involving
$B_{2}$, together with ones including $\mu$-term corrections compared
to numerical data for the $B_{2}$-vacuum and $B_{2}-B_{1}$ matrix
elements at $\xi=2/7$. }

\end{figure}

In order to do this, however, formula (\ref{eq:genffrelation}) must
be continued to complex rapidities, which requires matching the phases
of the matrix elements on the two sides. This is not difficult to
do by observing that the TCSA matrix elements are all real%
\footnote{This is the case because the TCSA representation of the Hamiltonian
turns out to be a real and symmetric matrix, due to the fact that
the matrix elements of operators $\mathrm{e}^{ia\beta\Phi}$ are all
real. As a results, all eigenvectors are also real and therefore all
the matrix elements computed from TCSA are real as well.%
}, while the phase of the minimal form factor function $f_{\xi}(\theta)$
(\ref{eq:brminff}) is just half the phase of the two-particle $S$-matrix
$S_{B_{1}B_{1}}(\theta)$. Using the results of \cite{Pozsgay:2008bf}
one obtains\[
\langle0\vert\mathcal{O}(0,0)\vert\{I\}\rangle_{2,L}=\pm\frac{\sqrt{S_{B_{1}B_{1}}(-2i\tilde{u})}F_{11}^{\mathcal{O}}(\tilde{\theta}+i\tilde{u},\tilde{\theta}-i\tilde{u})}{\sqrt{\rho_{11}(\tilde{\theta}+i\tilde{u},\tilde{\theta}-i\tilde{u})}}+\dots\]
where $\tilde{\theta}$, $\tilde{u}$ is the solution of (\ref{eq:b2_b1b1_by})
with the correct momentum, ie. \[
m_{1}L\sinh(\tilde{\theta}+i\tilde{u})+m_{1}L\sinh(\tilde{\theta}-i\tilde{u})=2\pi I\]
and the dots denote further (and generally much smaller) exponential
corrections; the $\pm$ sign accounts for the remaining phase ambiguity
from the square root. One can similarly evaluate $B_{1}-B_{2}$ matrix
elements using\begin{align*}
\left|_{1}\langle\{I'\}\vert\mathcal{O}(0,0)\vert\{I\}\rangle_{2,L}\right| & =\frac{\left|\sqrt{S_{B_{1}B_{1}}(-2i\tilde{u})}F_{111}^{\mathcal{O}}(i\pi+\tilde{\theta}',\tilde{\theta}+i\tilde{u},\tilde{\theta}-i\tilde{u})\right|}{\sqrt{m_{1}L\cosh\tilde{\theta}'\,\rho_{11}(\tilde{\theta}+i\tilde{u},\tilde{\theta}-i\tilde{u})}}+\dots\end{align*}
where the absolute values are necessary because we have not compensated
for the phase difference due to the presence of the additional $B_{1}$.
We observed substantial improvement over the naive matrix elements
(\ref{eq:genffrelation}) that neglect the $\mu$-term contributions;
some of our data are demonstrated in figure \ref{fig:mu-terms}. The
effect is very dramatic for the $B_{1}-B_{2}$ case, the naive prediction
drastically disagrees with the numerical results, while the inclusion
of the $\mu$-terms leads to excellent agreement. This is in fact
a bit unexpected, as the value for the exponent for $\xi=2/7$ is
\[
\mu L=\sqrt{m_{1}^{2}-\frac{m_{2}^{2}}{4}}L\approx0.37651\, l\qquad,\qquad l=ML\]
whose coefficient is not particularly small. In the original example
in \cite{Pozsgay:2008bf} one of the particles was loosely bound,
resulting in a very small value for $\mu$, which in turn led to large
finite size corrections from the particle splitting up in finite volume.
This is not case for $B_{2}$ considered as a bound state of two $B_{1}$s
at the particular value of the coupling $\xi$ in question. However,
the full behaviour is determined not only by the exponent, but also
by the detailed behaviour of the $S$-matrix amplitude and the form
factor near the pole; it is the latter that enhances the correction
in the $B_{1}-B_{2}$ case. We return to this observation and its
implications in the conclusion.

\subsection{\label{sub:Diagonal--matrix} Diagonal $B_{1}B_{1}-B_{1}B_{1}$ matrix
elements}

\begin{figure}
\begin{centering}
\psfrag{Volume}{$l$}
\psfrag{formfactor}{$|{}_{11}\langle\{I,-I\}|\mathcal{O}|\{I,-I\}\rangle_{11,L}|$}
\psfrag{ffb1b1s12Eb1b1s12DisconTCSA}{$I=1/2$, TCSA}
\psfrag{ffb1b1s12Eb1b1s12DisconTHEORY}{$I=1/2$, theory}
\psfrag{ffb1b1s32Eb1b1s32DisconTCSA}{$I=3/2$, TCSA}
\psfrag{ffb1b1s32Eb1b1s32DisconTHEORY}{$I=3/2$, theory}\includegraphics[bb=0bp 0bp 593bp 364bp,scale=0.7]{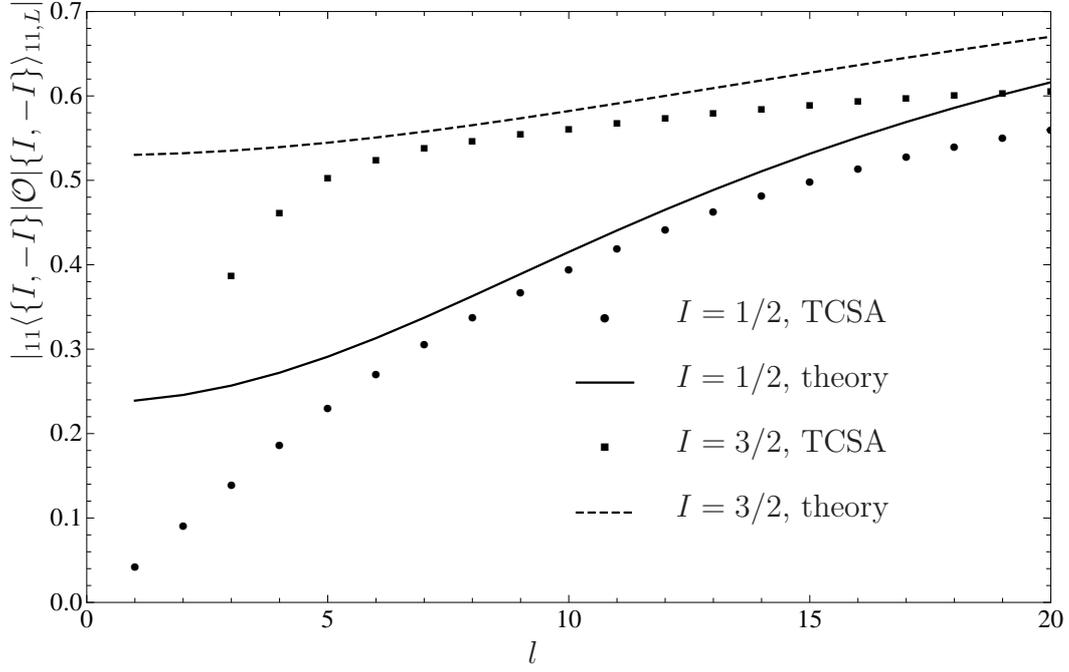}\\
\includegraphics[scale=0.7]{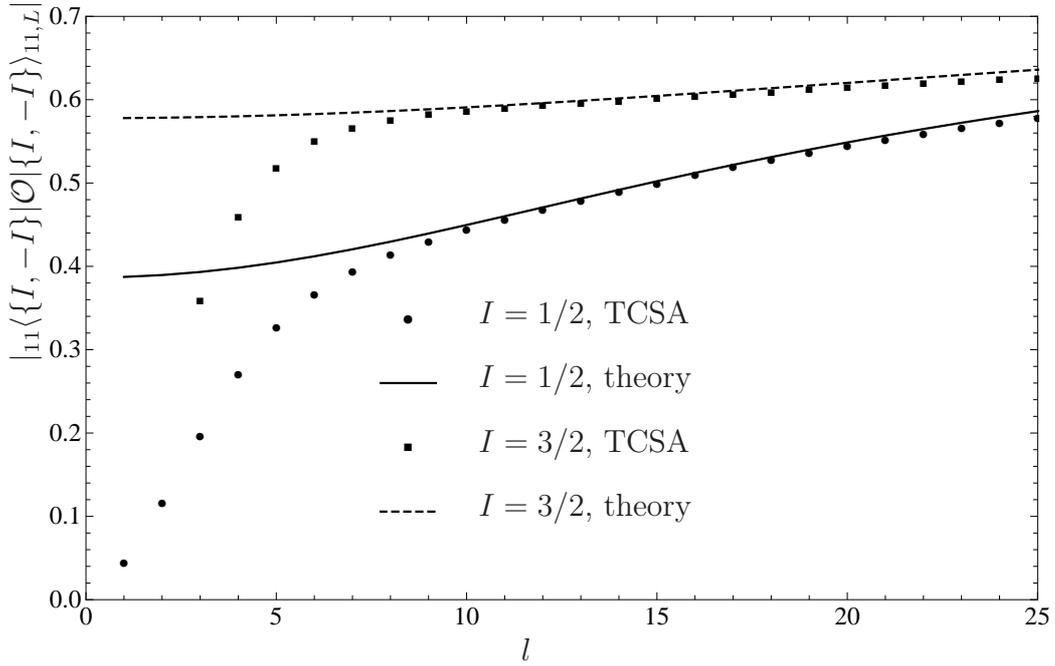}
\par\end{centering}

\caption{\label{fig:Diagonal-B1B1B1B1}Diagonal $B_{1}B_{1}-B_{1}B_{1}$ matrix
elements for $\xi=2/7$ and $\xi=50/311$.}

\end{figure}

For the diagonal matrix element we observe significant difference
between the numerical results and the theoretical predictions as shown
in figure \ref{fig:Diagonal-B1B1B1B1}. It is apparent from the data
that agreement is better for lower values of $\xi$. For large volumes,
this is easily explained by the better convergence of TCSA since the
dominant source of errors in the large volume data is the Hilbert
space truncation, and improvement of truncation errors with decreasing
$\xi$ was observed in all the data we analyzed. 

However, this cannot be the origin the deviation in the medium range
of volume ($5\lesssim l\lesssim15$). This deviation also decreases
for smaller $\xi$, and is likely to have its origin in a $\mu$-term
correction. Evaluating such a $\mu$-term requires modeling $B_{1}$
as a soliton-antisoliton bound state, which we are not able to do
at present as it necessitates numerically handling eight-particle
solitonic form factors. As for the variation of the $\mu$-term with
$\xi$, analogous behaviour was observed for the $B_{2}$ case, where
we also have analytic confirmation of the numerical results.

\subsection{Higher breathers}

We also tested numerous other form factors including higher breathers
(among them $B_{2}-B_{1}B_{1}$, $B_{2}$-$B_{2}$, $B_{1}-B_{3}$,
$B_{2}-B_{3}$ and $B_{3}-B_{3}$ matrix elements), and in all cases
found good agreement between theoretical predictions and TCSA data.
These tests cover all multi-$B_{1}$ form factors (\ref{eq:b1ffs})
up to $6$ particles (taking into account that $B_{n}$ is obtained
as a fusion of $n$ first breathers, as described in Appendix A of
\cite{Takacs:2009fu}).

\section{Soliton form factors in finite volume}

\subsection{Finite volume form factors in non-diagonal theories}

We can follow the lines of the reasoning in \cite{Pozsgay:2007kn}
that lead to the formula for the finite volume form factors. We recall
the main outlines of the arguments, making modifications where necessary.
We consider the spectral representation of the Euclidean two-point
function\begin{eqnarray}
\langle\mathcal{O}(\bar{x})\mathcal{O}'(0,0)\rangle & = & \sum_{n=0}^{\infty}\sum_{i_{1}\dots i_{n}}\left(\prod_{k=1}^{n}\int_{-\infty}^{\infty}\frac{d\theta_{k}}{2\pi}\right)F_{i_{1}\dots i_{n}}^{\mathcal{O}}(\theta_{1},\theta_{2},\dots,\theta_{n})\times\nonumber \\
 &  & \quad F_{i_{1}\dots i_{n}}^{\mathcal{O}'}(\theta_{1},\theta_{2},\dots,\theta_{n})^{+}\exp\left(-R\sum_{k=1}^{n}m_{i_{k}}\cosh\theta_{k}\right)\label{eq:spectralrepr}\end{eqnarray}
where\[
F_{i_{1}\dots i_{n}}^{\mathcal{O}'}(\theta_{1},\theta_{2},\dots,\theta_{n})^{+}=\,_{i_{1}\dots i_{n}}\langle\theta_{1},\dots,\theta_{n}|\mathcal{O}'(0,0)|0\rangle=F_{i_{1}\dots i_{n}}^{\mathcal{O}'}(\theta_{1}+i\pi,\theta_{2}+i\pi,\dots,\theta_{n}+i\pi)\]
(which is just the complex conjugate of $F_{n}^{\mathcal{O}'}$ for
unitary theories) and $R=\sqrt{\tau^{2}+x^{2}}$ is the length of
the Euclidean separation vector $\bar{x}=(\tau,x)$. 

In finite volume $L$, the space of states can still be labeled by
multi-particle states but the momenta (and therefore the rapidities)
are quantized. Denoting the quantum numbers $I_{1},\dots,I_{n}$ the
two-point function of the same local operator can be written as\begin{eqnarray}
\langle\mathcal{O}(\tau,0)\mathcal{O}'(0,0)\rangle_{L} & = & \sum_{n=0}^{\infty}\sum_{r}\sum_{I_{1}\dots I_{n}}\langle0\vert\mathcal{O}(0,0)\vert\{I_{1},I_{2},\dots,I_{n}\}\rangle_{L}^{(r)}\times\label{eq:finitevolspectralrepr}\\
 &  & \quad^{(r)}\langle\{I_{1},I_{2},\dots,I_{n}\}|\mathcal{O}'(0,0)|0\rangle_{L}\exp\left(-\tau\sum_{k=1}^{n}m_{i_{k}}\cosh\theta_{k}\right)\nonumber \end{eqnarray}
where the index $r$ enumerates the eigenvector of the $n$-particle
transfer matrix (as an example cf. Appendix A where these are constructed
for sine-Gordon solitons); these are usually rapidity dependent linear
combinations in the space of particle multiplet indices $i_{1}\dots i_{n}$.
We can assume that the wave function amplitudes (\ref{eq:internalwavefunction})
of these states are normalized and form a complete basis:\begin{align*}
\sum_{i_{1}\dots i_{n}}\Psi_{i_{1}\dots i_{n}}^{(r)}\left(\left\{ \theta_{k}\right\} \right)\Psi_{i_{1}\dots i_{n}}^{(s)}\left(\left\{ \theta_{k}\right\} \right)^{*} & =\delta_{rs}\\
\sum_{r}\Psi_{i_{1}\dots i_{n}}^{(r)}\left(\left\{ \theta_{k}\right\} \right)\Psi_{j_{1}\dots j_{n}}^{(r)}\left(\left\{ \theta_{k}\right\} \right)^{*} & =\delta_{i_{1}j_{1}}\dots\delta_{i_{n}j_{n}}\end{align*}
In writing (\ref{eq:finitevolspectralrepr}) we also supposed that
the finite volume multi-particle states $\vert\{I_{1},I_{2},\dots,I_{n}\}\rangle_{r,L}$
are orthonormal and for simplicity restricted the formula to separation
in Euclidean time $\tau$ only. The index $L$ signals that the matrix
element is evaluated in finite volume $L$. Using the finite volume
expansion developed by Lüscher in \cite{Luscher:1985dn} one can easily
see that \begin{equation}
\langle\mathcal{O}(\tau,0)\mathcal{O}'(0)\rangle-\langle\mathcal{O}(\tau,0)\mathcal{O}'(0)\rangle_{L}\sim O(\mathrm{e}^{-\mu L})\label{eq:corrfinvoldiff}\end{equation}
where $\mu$ is some characteristic mass scale. 

We can now rewrite the infinite volume correlator in the basis of
multi-particle transfer matrix eigenstates\begin{eqnarray*}
\langle\mathcal{O}(\tau,0)\mathcal{O}'(0,0)\rangle & = & \sum_{n=0}^{\infty}\sum_{r}\left(\prod_{k=1}^{n}\int_{-\infty}^{\infty}\frac{d\theta_{k}}{2\pi}\right)F^{\mathcal{O}}(\theta_{1},\dots,\theta_{n})^{(r)}\times\\
 &  & \quad F^{\mathcal{O}'}(\theta_{1},\dots,\theta_{n})^{(r)+}\exp\left(-\tau\sum m_{i_{k}}\cosh\theta_{k}\right)\end{eqnarray*}
where \[
F^{\mathcal{O}}(\theta_{1},\dots,\theta_{n})^{(r)}=\sum_{i_{1}\dots i_{n}}F_{i_{1}\dots i_{n}}^{\mathcal{O}}(\theta_{1},\dots,\theta_{n})\Psi_{i_{1}\dots i_{n}}^{(r)}\left(\left\{ \theta_{k}\right\} \right)\]
The remainder of the argument follows the lines of the paper \cite{Pozsgay:2007kn}.
Essentially, we compare the discrete sum with the integral, and realize
that up to exponentially small terms in $L$ the relation is given
by the Jacobian of the mapping between the quantum number and the
rapidity space, i.e. the density of states. We obtain \begin{equation}
\left|\langle0\vert\mathcal{O}(0,0)\vert\{I_{1},\dots,I_{n}\}\rangle_{L}^{(r)}\right|=\left|\frac{F^{\mathcal{O}}(\tilde{\theta}_{1},\dots,\tilde{\theta}_{n})^{(r)}}{\sqrt{\rho^{(r)}(\tilde{\theta}_{1},\dots,\tilde{\theta}_{n})}}\right|+O(\mathrm{e}^{-\mu'L})\label{eq:ffrelation}\end{equation}
where%
\footnote{We note that $\mu'$ is not necessarily the same scale as in (\ref{eq:corrfinvoldiff})
and its value depends on the spectrum and bound state fusion angles
of the model.%
}\[
\rho^{(r)}(\theta_{1},\dots,\theta_{n})\]
is the density of states with internal wave vector $\Psi^{(r)}$ as
given in (\ref{eq:nondiagjacobi}), and $\tilde{\theta}_{k}$ are
the solutions of the Bethe-Yang equations (\ref{eq:nondiagby}) corresponding
to the state with the specified quantum numbers $I_{1},\dots,I_{n}$
at the given volume $L$. 

This can be easily generalized to matrix elements with no disconnected
pieces:\begin{eqnarray}
 &  & \left|\,^{(s)}\langle\{I_{1}',\dots,I_{M}'\}\vert\mathcal{O}(0,0)\vert\{I_{1},\dots,I_{N}\}\rangle_{L}^{(r)}\right|=\nonumber \\
 &  & \qquad\left|\frac{{\displaystyle F^{\mathcal{O}}(\tilde{\theta}_{M},\dots,\tilde{\theta}_{1}|\tilde{\theta}_{1},\dots,\tilde{\theta}_{N})^{(s,r)}}}{\sqrt{\rho^{(r)}(\tilde{\theta}_{1},\dots,\tilde{\theta}_{N})\rho^{(s)}(\tilde{\theta}_{1}',\dots,\tilde{\theta}_{M}')}}\right|+O(\mathrm{e}^{-\mu'L})\label{eq:nondiag_genffrelation}\end{eqnarray}
where\begin{eqnarray*}
 &  & F^{\mathcal{O}}(\tilde{\theta}_{M}',\dots,\tilde{\theta}_{1}'|\tilde{\theta}_{1},\dots,\tilde{\theta}_{N})^{(s,r)}\\
 &  & =\sum_{j_{1}\dots j_{M}}\sum_{i_{1}\dots i_{N}}\Psi_{j_{1}\dots j_{M}}^{(s)}\left(\left\{ \tilde{\theta}_{k}'\right\} \right)^{*}F_{\bar{j}_{M}\dots\bar{j}_{1}i_{1}\dots i_{N}}^{\mathcal{O}}(\tilde{\theta}_{M}'+i\pi,\dots,\tilde{\theta}_{1}'+i\pi,\tilde{\theta}_{1},\dots,\tilde{\theta}_{N})\Psi_{i_{1}\dots i_{N}}^{(r)}\left(\left\{ \tilde{\theta}_{k}\right\} \right)\end{eqnarray*}
where the bar denotes the antiparticle.

This can be easily applied to soliton-antisoliton states. The algebraic
Bethe Ansatz gives two eigenvectors in this subspace (cf. \ref{sub:The-two-particle-case}):\begin{align*}
B(\lambda_{1})\Omega & \propto\begin{cases}
v_{+-}+v_{-+} & \quad\mbox{for }\lambda_{1}=\frac{\theta_{1}+\theta_{2}}{2}+i\frac{\pi}{2}\\
v_{+-}-v_{-+} & \quad\mbox{for }\lambda_{1}=\frac{\theta_{1}+\theta_{2}}{2}+i\frac{(1+\xi)\pi}{2}\end{cases}\end{align*}
i.e.\begin{align*}
\Psi^{(+)} & =\frac{1}{\sqrt{2}}(0,+1,+1,0)\\
\Psi^{(-)} & =\frac{1}{\sqrt{2}}(0,+1,-1,0)\end{align*}
Since the value of the magnonic variable $\lambda_{1}$ is explicitly
known, it can be eliminated from the Bethe-Yang equations, resulting
in the following quantization conditions for the soliton-antisoliton
pair:\begin{align}
Q_{1}^{(\pm)}(\theta_{1},\theta_{2}) & =ML\sinh\theta_{1}-i\log\mathcal{S}_{\pm}(\theta_{1}-\theta_{2})=2\pi I_{1}\nonumber \\
Q_{2}^{(\pm)}(\theta_{1},\theta_{2}) & =ML\sinh\theta_{2}-i\log\mathcal{S}_{\pm}(\theta_{2}-\theta_{1})=2\pi I_{2}\label{eq:ssbarby}\end{align}
where $\mathcal{S}_{\pm}$ are the eigenvalues of the two-particle
$S$-matrix given in (\ref{eq:tps_evals}) and the $\pm$ distinguishes
the two states. It is also easy to eliminate the magnon $\lambda_{1}$
from the density of states to obtain\[
\rho^{(\pm)}(\theta_{1},\theta_{2})=\left|\begin{array}{cc}
\frac{\partial Q_{1}^{(\pm)}}{\partial\theta_{1}} & \frac{\partial Q_{1}^{(\pm)}}{\partial\theta_{2}}\\
\frac{\partial Q_{2}^{(\pm)}}{\partial\theta_{1}} & \frac{\partial Q_{2}^{(\pm)}}{\partial\theta_{2}}\end{array}\right|\]
From (\ref{eq:ffrelation}) we obtain\begin{equation}
\left|\,\langle0\vert\mathcal{O}(0,0)\vert\{I_{1},I_{2}\}\rangle_{L}^{(\pm)}\right|=\frac{\left|F^{\pm}(\tilde{\theta}_{1}-\tilde{\theta}_{2})\right|}{\sqrt{\rho^{(\pm)}(\tilde{\theta}_{1},\tilde{\theta}_{2})}}+O(\mathrm{e}^{-\mu L})\label{eq:ssbar_finitevol}\end{equation}
where \[
F^{\pm}(\theta)=\frac{1}{\sqrt{2}}\left(F_{+-}^{\beta}(\theta)\pm F_{-+}^{\beta}(\theta)\right)\]
in terms of (\ref{eq:twossff}) and $\tilde{\theta}_{1,2}$ are the
solutions of (\ref{eq:ssbarby}) at the given volume $L$ with quantum
numbers $I_{1,2}$.

At the moment, there is no general theoretical result for matrix elements
with disconnected pieces (e.g. diagonal ones). However, for one-soliton
matrix elements one can easily write down the appropriate generalization
of eqn. (\ref{eq:br_diagff}):\begin{equation}
\,_{+}\langle\{I\}\vert\mathcal{O}(0,0)\vert\{I\}\rangle_{+}=\frac{F_{-+}(-i\pi)}{ML\cosh\tilde{\theta}}+\mathcal{G}_{1}(\beta)\label{eq:ss_diagff}\end{equation}
where $F_{-+}$ is given in (\ref{eq:twossff}) and the rapidity $\tilde{\theta}$
is quantized as \[
ML\sinh\tilde{\theta}=2\pi I\]

\subsection{Numerical verification of solitonic matrix elements}

\begin{figure}
\begin{centering}
\psfrag{Volume}{$l$}
\psfrag{formfactor}{$|\langle 0|\mathcal{O}|{I_1,I_2}\rangle^{(+)}_L|$}
\psfrag{sam32ffSim}{$I_1=-I_2=3/2$, TCSA}
\psfrag{sam32ffTheory}{$I_1=-I_2=3/2$, theory}
\psfrag{sam52ffSim}{$I_1=-I_2=5/2$, TCSA}
\psfrag{sam52ffTheory}{$I_1=-I_2=5/2$, theory}\includegraphics[scale=0.7]{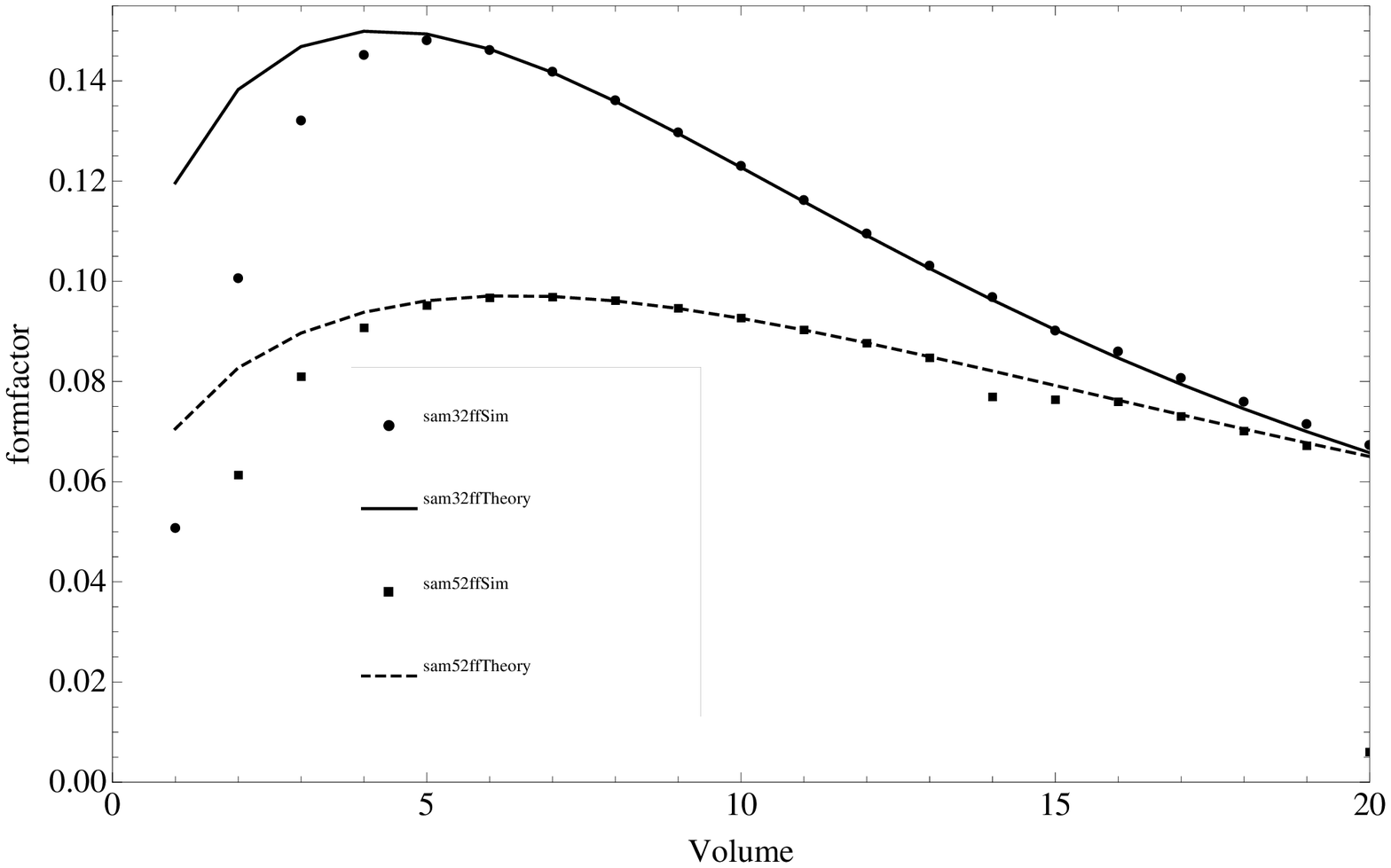}\\
\psfrag{Volume}{$l$}
\psfrag{formfactor}{|$\langle 0|\mathcal{O}|{I_1,I_2}\rangle^{(-)}_L$|}
\psfrag{sap1ffSim}{$I_1=-I_2=1$, TCSA}
\psfrag{sap1ffTheory}{$I_1=-I_2=1$, theory}\includegraphics[scale=0.7]{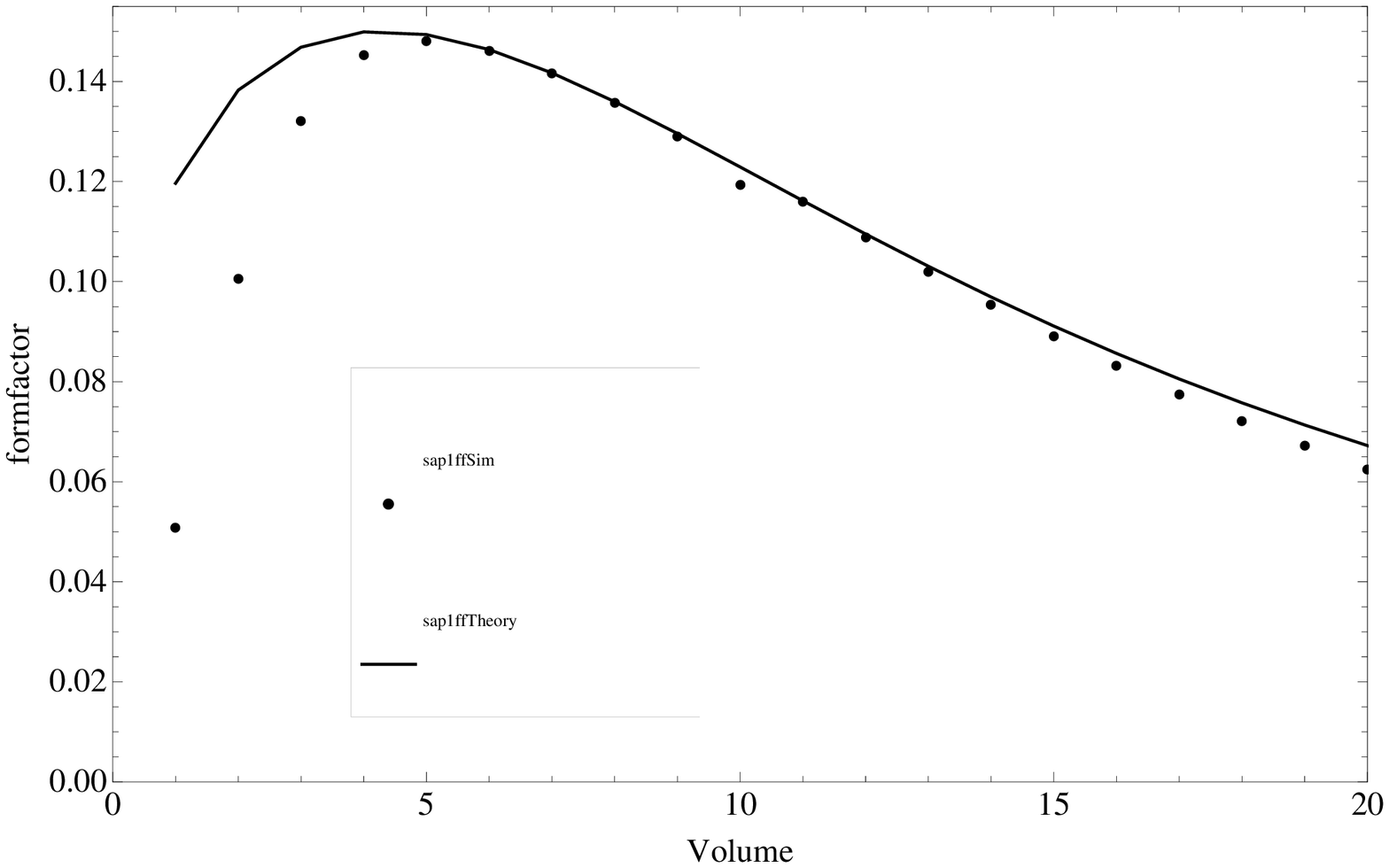}
\par\end{centering}

\caption{\label{fig:Vacuum--soliton-antisoliton-form-factors}Vacuum--soliton-antisoliton
form factors at $\xi=2/7$. The individual deviations are caused by
the proximity of level crossings.}
\end{figure}

\begin{figure}
\begin{centering}
\psfrag{Rapidity}{$l$}
\psfrag{formfactor}{|$\langle {I'}|\mathcal{O}|{I}\rangle_L$|}
\psfrag{ffss0Ess0DisconTCSA}{$I=I'=0$, TCSA}
\psfrag{ffss0Ess0DisconTHEORY}{$I=I'=0$, theory}
\psfrag{ffss1Ess1DisconTCSA}{$I=I'=1$, TCSA}
\psfrag{ffss1Ess1DisconTHEORY}{$I=I'=1$, theory}
\psfrag{ffss1Ess0TCSA}{$I=0$, $I'=1$, TCSA}
\psfrag{ffss1Ess0THEORY}{$I=0$, $I'=1$, theory}\includegraphics[scale=0.7]{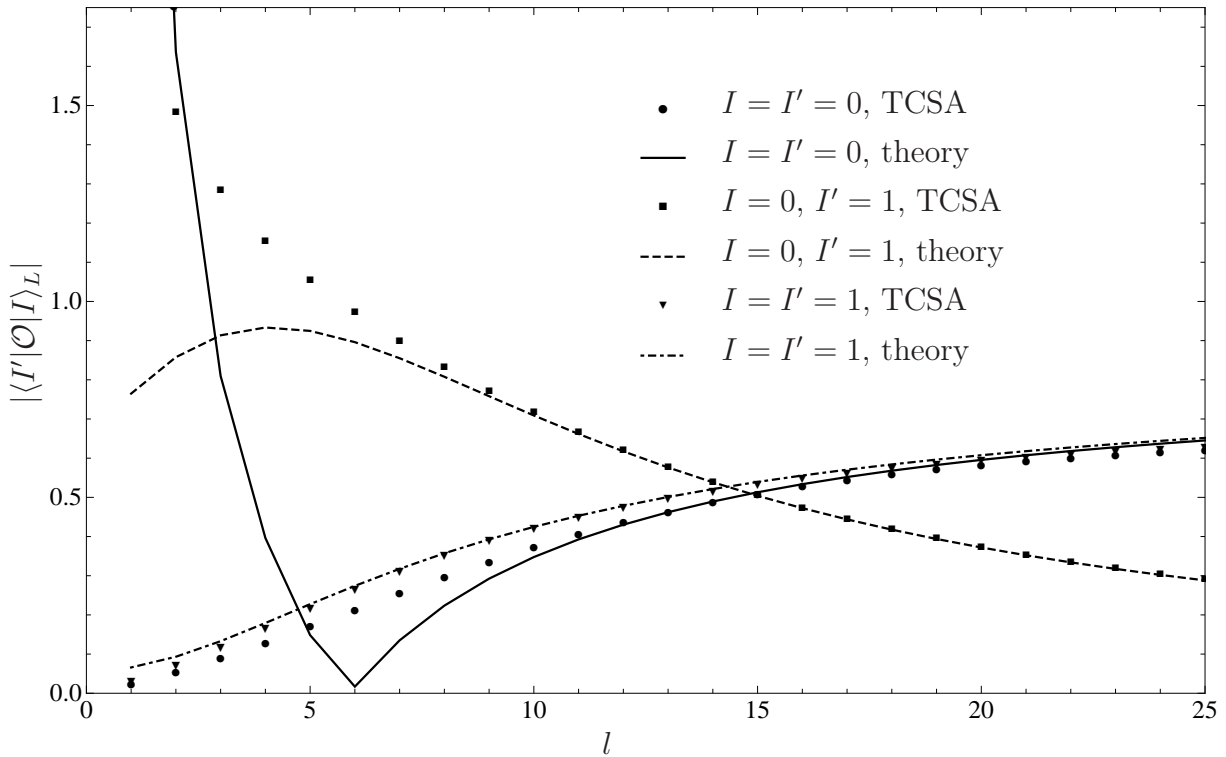}
\par\end{centering}

\caption{\label{fig:Soliton--soliton-form-factors:}Soliton--soliton form factors:
off-diagonal and diagonal case at $\xi=50/239$. The cusp in the continuous
line is an artifact caused by taking the absolute value.}
\end{figure}

The formulae derived in the previous subsection show an excellent
agreement with TCSA. Figure \ref{fig:Vacuum--soliton-antisoliton-form-factors}
demonstrates this for the matrix element between the vacuum and the
(symmetric or antisymmetric) soliton-antisoliton two-particle states
as given in eqn. (\ref{eq:ssbar_finitevol}). We can also study one-soliton--one-soliton
matrix elements. For the non-diagonal ones we can use the $N=M=1$
case of (\ref{eq:nondiag_genffrelation}) while for the diagonal ones
it turns out we must slightly modify (\ref{eq:ss_diagff}) by changing
the sign of the $F_{-+}$ term:\begin{equation}
\,_{+}\langle\{I\}\vert\mathcal{O}(0,0)\vert\{I\}\rangle_{+}=-\frac{F_{-+}(-i\pi)}{ML\cosh\tilde{\theta}}+\mathcal{G}_{1}(\beta)\label{eq:ss_diagff-1}\end{equation}
It is not yet clear to us what is the reason behind this, but, as
shown in figure \ref{fig:Soliton--soliton-form-factors:} this produces
an excellent agreement with the TCSA data. It is possible that the
bootstrap solution (\ref{eq:twossff}) is off by a sign%
\footnote{This would not be detected by checking off-diagonal matrix elements,
since due to phase differences these compare only the absolute values
according to (\ref{eq:nondiag_genffrelation}).%
}, or that the sign here is related in some way to the crossing of
a soliton. 

\section{Conclusions and outlook}

In this work we investigated the comparison between form factors of
sine-Gordon theory obtained from the bootstrap and finite volume matrix
elements given by the truncated conformal space approach. Generally
speaking, we found an excellent agreement between the theoretical
predictions and the numerical results. We also generalized the formalism
developed in \cite{Pozsgay:2007kn,Pozsgay:2007gx} to non-diagonal
scattering theory in order to include solitonic matrix elements in
the test. 

In this latter respect, however, several problems are still open.
One of them is to find a suitable formula for matrix elements with
disconnected contributions in the case of non-diagonal scattering
(at present we only have the simplest, one-particle--one-particle
case covered). The other, more technical problem is to find a suitable
framework for the numerical evaluation of the multi-soliton form factors;
until then, the full strength of the algebraic Bethe Ansatz formalism
presented in Appendix \ref{sec:Algebraic-Bethe-Ansatz} cannot be
utilized. In addition, the solution of the problem in the case of
diagonal $B_{1}B_{1}-B_{1}B_{1}$ matrix elements, described in subsection
\ref{sub:Diagonal--matrix}, also hinges on the ability of handling
such matrix elements. A further issue is the presence of the minus
sign in (\ref{eq:ss_diagff-1}) which requires a detailed look into
the soliton form factor bootstrap, which is out of the scope of present
work.

Another interesting issue is the role played by the so-called $\mu$-terms,
which are exponential corrections to the volume dependence of the
matrix elements resulting from particle fusions (in other words, three-particle
couplings between on-shell particle states). The exponential decay
of these corrections depends only on the mass spectrum; however, their
strength is also determined by appropriate form factors, as illustrated
in subsection \ref{sub:mu-terms-for-b2} by the fact that for the
same $B_{2}$ state (hence the same $\mu$-term kinematics), the significance
of the corrections depends very much on the matrix element the state
is involved in.

In this work we generally took the stance of using the bootstrap results
to predict the numerical finite volume results. However, one could
take the opposite route and extract field theoretically interesting
matrix elements from the finite volume data (which we also did for
cases involving two-particle matrix elements). This is common in lattice
field theory and was advocated by Lellouch and Lüscher in \cite{Lellouch:2000pv}
as a way to circumvent the so-called Maiani-Testa no-go theorem \cite{Maiani:1990ca}.
We have previously used a similar approach to determine resonance
parameters from finite volume spectra \cite{Pozsgay:2006wb}. 

The $\mu$-term results have an implication for such studies, which
aim to determine matrix elements (e.g. related to weak interaction)
from finite volume data (whether using lattice Monte-Carlo or some
other numerical method). The important lesson is that even if the
appropriate exponential factor $\mu$ (as determined by the mass spectrum)
is not small and therefore volume suppression is expected to be strong,
the $\mu$-term can still have a sizable effect at the values of volume
the applied numerical method (whether TCSA or lattice) can reach.
In addition, this term could vary strongly between different matrix
elements involving the same state, thereby producing very large systematic
errors when extracting matrix elements of local operators from numerical
simulations (indeed, the corrections could of the order of 100\% as
illustrated by the $B_{2}-B_{1}$ matrix element plotted in figure
\ref{fig:mu-terms}).

\subsubsection*{Acknowledgments}

GT is grateful to B. Pozsgay for useful discussions. This work was
partially supported by the Hungarian OTKA grants K75172 and K81461.

\appendix
\makeatletter 
\renewcommand{\theequation}{\hbox{\normalsize\Alph{section}.\arabic{equation}}} \@addtoreset{equation}{section} \renewcommand{\thefigure}{\hbox{\normalsize\Alph{section}.\arabic{figure}}} \@addtoreset{figure}{section} \renewcommand{\thetable}{\hbox{\normalsize\Alph{section}.\arabic{table}}} \@addtoreset{table}{section} \makeatother

\section{\label{sec:Algebraic-Bethe-Ansatz} Algebraic Bethe Ansatz for multi-soliton
states}

\subsection{The sine-Gordon soliton Bethe-Yang equations and the multi-particle
transfer matrix}

Our starting object is the $N$-particle monodromy matrix that can
be written as\[
\mathcal{M}\left(\lambda|\left\{ \theta_{1},\dots,\theta_{N}\right\} \right)_{a,\, i_{1}\dots i_{N}}^{b,\, j_{1}\dots j_{N}}=\mathcal{S}_{ai_{1}}^{c_{1}j_{1}}(\lambda-\theta_{1})\mathcal{S}_{c_{1}i_{2}}^{c_{2}j_{2}}(\lambda-\theta_{2})\dots\mathcal{S}_{c_{N-1}i_{N}}^{bj_{N}}(\lambda-\theta_{N})\]
or as a $2\times2$ matrix in the $a,b$ indices\[
\mathcal{M}\left(\lambda|\left\{ \theta_{k}\right\} \right)=\left(\begin{array}{cc}
A\left(\lambda|\left\{ \theta_{k}\right\} \right) & B\left(\lambda|\left\{ \theta_{k}\right\} \right)\\
C\left(\lambda|\left\{ \theta_{k}\right\} \right) & D\left(\lambda|\left\{ \theta_{k}\right\} \right)\end{array}\right)\]
The operators $A$, $B$, $C$ and $D$ act in the $2^{N}$-dimensional
{}``isospin'' space (the space spanned by the $N$ soliton doublets)\[
\mathcal{V}_{N}=\bigotimes_{k=1}^{n}\mathbb{C}^{2}\]
which is spanned by the basis\[
v_{i_{1}\dots i_{N}}\quad,\quad i_{k}=\pm\]
One can introduce the vector\[
\Omega=v_{++\dots+}\]
corresponding to all solitons being positively charged. The general
transfer matrix\[
\mathcal{T}\left(\lambda|\left\{ \theta_{k}\right\} \right)=\mathcal{M}\left(\lambda|\left\{ \theta_{k}\right\} \right)_{a}^{a}=A\left(\lambda|\left\{ \theta_{k}\right\} \right)+D\left(\lambda|\left\{ \theta_{k}\right\} \right)\]
gives rise to the specified transfer matrices\[
\tau_{j}\left(\left\{ \theta_{k}\right\} \right)=\mathcal{T}\left(\lambda=\theta_{j}|\left\{ \theta_{k}\right\} \right)\]
In terms of these the $n$-particle quantization relations, ie. the
Bethe-Yang equations on the circle can be written as \cite{Takacs:1998df}
\[
\mathrm{e}^{iML\sinh\theta_{j}}\tau_{j}\left(\left\{ \theta_{k}\right\} \right)\Psi\left(\left\{ \theta_{k}\right\} \right)=\Psi\left(\left\{ \theta_{k}\right\} \right)\]
where $\Psi\left(\left\{ \theta_{k}\right\} \right)\in\mathcal{V}_{N}$
is the wave-function amplitude vector.

\subsection{The Algebraic Bethe Ansatz}

We use the well-established machinery of the algebraic Bethe Ansatz
(for a pedagogical introduction see \cite{Faddeev:1996iy}). The sine-Gordon
$S$-matrix satisfies the Yang-Baxter equations\[
\mathcal{S}_{k_{2}k_{3}}^{j_{2}j_{3}}(\theta_{23})\mathcal{S}_{k_{1}i_{3}}^{j_{1}k_{3}}(\theta_{13})\mathcal{S}_{i_{1}i_{2}}^{k_{1}k_{2}}(\theta_{12})=\mathcal{S}_{k_{1}k_{2}}^{j_{1}j_{2}}(\theta_{12})\mathcal{S}_{i_{1}k_{3}}^{k_{1}j_{3}}(\theta_{13})\mathcal{S}_{i_{2}i_{3}}^{k_{2}k_{3}}(\theta_{23})\]
where $\theta_{ij}=\theta_{i}-\theta_{j}$. As a result, the monodromy
matrix satisfies the relations\[
\mathcal{S}_{b_{1}b_{2}}^{c_{1}c_{2}}(\lambda-\mu)\mathcal{M}\left(\lambda|\left\{ \theta_{k}\right\} \right)_{a_{1}}^{b_{1}}\mathcal{M}\left(\mu|\left\{ \theta_{k}\right\} \right)_{a_{2}}^{b_{2}}=\mathcal{M}\left(\mu|\left\{ \theta_{k}\right\} \right)_{d_{1}}^{c_{1}}\mathcal{M}\left(\lambda|\left\{ \theta_{k}\right\} \right)_{d_{2}}^{c_{2}}\mathcal{S}_{c_{1}c_{2}}^{d_{1}d_{2}}(\lambda-\mu)\]
Therefore the transfer matrices form a commuting family\[
\mathcal{T}\left(\lambda|\left\{ \theta_{k}\right\} \right)\mathcal{T}\left(\mu|\left\{ \theta_{k}\right\} \right)=\mathcal{T}\left(\mu|\left\{ \theta_{k}\right\} \right)\mathcal{T}\left(\lambda|\left\{ \theta_{k}\right\} \right)\]
and can be diagonalized simultaneously with common eigenvectors for
all $\lambda$:\[
\mathcal{T}\left(\lambda|\left\{ \theta_{k}\right\} \right)\psi_{\alpha}\left(\left\{ \theta_{k}\right\} \right)=\Lambda_{\alpha}\left(\lambda|\left\{ \theta_{k}\right\} \right)\psi_{\alpha}\left(\left\{ \theta_{k}\right\} \right)\]
where $\Lambda_{\alpha}\left(\lambda|\left\{ \theta_{k}\right\} \right)\in S^{1}$
are the eigenvalues, $\psi_{\alpha}\left(\left\{ \theta_{k}\right\} \right)\in\mathcal{V}_{n}$
are the eigenvectors and $\alpha=1\dots2^{N}$. The independent solutions
of the Bethe equations for the wave vector are then given by\[
\Psi\left(\left\{ \theta_{k}\right\} \right)=\psi_{\alpha}\left(\left\{ \theta_{k}\right\} \right)\]
provided the particle rapidities satisfy the Bethe-Yang equations\[
\mathrm{e}^{iML\sinh\theta_{j}}\Lambda_{\alpha}\left(\theta_{j}|\left\{ \theta_{k}\right\} \right)=1\qquad j=1\dots N\]
In this way we reduced the problem to finding the eigenvalues of the
transfer matrix and then solving a system of $N$ coupled scalar equations.

Note that the sine-Gordon scattering preserves the solitonic charge
$\mathcal{Q}$ and charge parity $\mathcal{C}$, which act on $\mathcal{V}_{N}$
as follows:\begin{eqnarray*}
\mathcal{Q}v_{i_{1}\dots i_{N}} & = & \left(\sum_{k=1}^{N}i_{k}\right)v_{i_{1}\dots i_{N}}\\
\mathcal{C}v_{i_{1}\dots i_{N}} & = & v_{\bar{i}_{1}\dots\bar{i}_{N}}\qquad\mbox{where }\bar{i}=-i\end{eqnarray*}
The transfer matrix $\mathcal{T}$ commutes with these operators.
In addition, $B$ and $C$ act as lowering and raising operators in
charge space\begin{eqnarray*}
\left[\mathcal{Q},B\left(\lambda|\left\{ \theta_{k}\right\} \right)\right] & = & -2B\left(\lambda|\left\{ \theta_{k}\right\} \right)\\
\left[\mathcal{Q},C\left(\lambda|\left\{ \theta_{k}\right\} \right)\right] & = & +2C\left(\lambda|\left\{ \theta_{k}\right\} \right)\end{eqnarray*}
We now look for the eigenvectors of the transfer matrix in the form\begin{equation}
\Psi\left(\left\{ \lambda_{s}\right\} |\left\{ \theta_{k}\right\} \right)=\mathcal{N}_{\Psi}\, B\left(\lambda_{1}|\left\{ \theta_{k}\right\} \right)\dots B\left(\lambda_{r}|\left\{ \theta_{k}\right\} \right)\Omega\label{eq:internalwavefunction}\end{equation}
where $\lambda_{s}$ are the so-called 'magnons' and $\mathcal{N}_{\Psi}$
is some normalization factor. They satisfy the eigenvalue equations\[
\mathcal{T}\left(\lambda|\left\{ \theta_{1},\dots,\theta_{n}\right\} \right)\Psi\left(\left\{ \lambda_{s}\right\} |\left\{ \theta_{k}\right\} \right)=\Lambda\left(\lambda,\left\{ \lambda_{s}\right\} |\left\{ \theta_{k}\right\} \right)\Psi\left(\left\{ \lambda_{s}\right\} |\left\{ \theta_{k}\right\} \right)\]
provided the algebraic Bethe Ansatz (ABA) equations hold\[
\prod_{k=1}^{N}a(\lambda_{j}-\theta_{k})=\prod_{k\neq j}^{r}\frac{a(\lambda_{j}-\lambda_{k})}{a(\lambda_{k}-\lambda_{j})}\qquad a(\theta)=\frac{1}{S_{T}(\theta,\xi)}=\frac{\sinh\left(\frac{i\pi-\theta}{\xi}\right)}{\sinh\left(\frac{\theta}{\xi}\right)}\]
The eigenvalue is given by\begin{eqnarray*}
\Lambda\left(\lambda,\left\{ \lambda_{s}\right\} |\left\{ \theta_{1},\dots,\theta_{N}\right\} \right) & = & \left(\prod_{k=1}^{r}S_{T}(\lambda_{k}-\lambda)^{-1}+\prod_{k=1}^{N}S_{T}(\lambda-\theta_{k},\xi)\prod_{k=1}^{r}S_{T}(\lambda-\lambda_{k})^{-1}\right)\\
 &  & \times\prod_{k=1}^{N}S_{0}(\lambda-\theta_{k},\xi)\end{eqnarray*}
Notice that $a(\theta)$ (and therefore the ABA equations) has the
natural periodicity \[
a(\theta+i\pi\xi)=a(\theta)\]

\subsection{\label{sub:The-two-particle-case} The two-particle case ($N=2$)}

The only interesting eigenvectors are the ones in the $\mathcal{Q}=0$
subspace, for which we can write the Ansatz\[
B(\lambda_{1})\Omega\]
The single ABA equation takes the form \[
a(\lambda_{1}-\theta_{1})a(\lambda_{1}-\theta_{2})=\frac{\sinh\left(\frac{i\pi-\lambda_{1}+\theta_{1}}{\xi}\right)}{\sinh\left(\frac{\lambda_{1}-\theta_{1}}{\xi}\right)}\frac{\sinh\left(\frac{i\pi-\lambda_{1}+\theta_{2}}{\xi}\right)}{\sinh\left(\frac{\lambda_{1}-\theta_{2}}{\xi}\right)}=1\]
and (up to periodicity) has two independent solutions\begin{eqnarray*}
(1) & : & \lambda_{1}=\frac{\theta_{1}+\theta_{2}}{2}+i\frac{\pi}{2}\\
(2) & : & \lambda_{1}=\frac{\theta_{1}+\theta_{2}}{2}+i\pi\frac{1+\xi}{2}\end{eqnarray*}
We can then evaluate the corresponding eigenvalues:\[
\Lambda\left(\lambda,\left\{ \lambda_{1}\right\} |\left\{ \theta_{1},\theta_{2}\right\} \right)=\left(\frac{1}{S_{T}(\lambda_{1}-\lambda)}+\frac{S_{T}(\lambda-\theta_{1})S_{T}(\lambda-\theta_{2})}{S_{T}(\lambda-\lambda_{1})}\right)S_{0}(\lambda-\theta_{1})S_{0}(\lambda-\theta_{2})\]
Putting $\lambda=\theta_{1}$ (to get the Bethe-Yang phase-shift for
the first particle) we find\begin{eqnarray*}
(1) & : & \Lambda\left(\theta_{1},\left\{ \lambda_{1}\right\} |\left\{ \theta_{1},\theta_{2}\right\} \right)=-\mathcal{S}_{+}(\theta_{1}-\theta_{2})\\
(2) & : & \Lambda\left(\theta_{1},\left\{ \lambda_{1}\right\} |\left\{ \theta_{1},\theta_{2}\right\} \right)=-\mathcal{S}_{-}(\theta_{1}-\theta_{2})\end{eqnarray*}
where \begin{equation}
\mathcal{S}_{\pm}(\theta)=\left(S_{T}(\theta)\pm S_{R}(\theta)\right)S_{0}(\theta)\label{eq:tps_evals}\end{equation}
are the eigenvalues of the two-particle S matrix in the neutral subspace.

\subsection{The four-particle case ($N=4$)}

Albeit we are not able to use them for detailed numerical comparison
yet, we briefly review the four-particle results to give a better
understanding of what follows. Suppose that the four rapidities are
ordered as\[
\theta_{1}<\theta_{2}<\theta_{3}<\theta_{4}\]

\subsubsection{$Q=2$ sector}

This subspace is four dimensional, and with the Ansatz\[
B(\lambda_{1})\Omega\]
there are two types of solutions:
\begin{itemize}
\item $\lambda_{1}=\mu+i\frac{\pi}{2}$\\
There are three such solutions, typically one with $\mu$ the between
$\theta_{2}$ and $\theta_{3}$, one around $\theta_{1}$ and another
one around $\theta_{4}$.
\item $\lambda_{1}=\mu+i(1+\xi)\frac{\pi}{2}$\\
There is a single solution, with $\mu$ the between $\theta_{2}$
and $\theta_{3}$.
\end{itemize}

\subsubsection{$Q=0$ sector}

There are $6$ eigenvectors here, which can be classified by their
parity under charge conjugation $\mathcal{C}$. The Ansatz is

\[
B(\lambda_{1})B(\lambda_{2})\Omega\]
and the ABA equations are\[
\prod_{k=1}^{4}S_{T}(\lambda_{1}-\theta_{k})=\frac{S_{T}(\lambda_{1}-\lambda_{2})}{S_{T}(\lambda_{2}-\lambda_{1})}\qquad\prod_{k=1}^{4}S_{T}(\lambda_{2}-\theta_{k})=\frac{S_{T}(\lambda_{2}-\lambda_{1})}{S_{T}(\lambda_{1}-\lambda_{2})}\]
For the three $\mathcal{C}=-1$ eigenvectors, the magnons take the
form\[
\lambda_{1}=\mu_{1}+\frac{i\pi}{2}\quad\lambda_{2}=\mu_{2}+\frac{i(1+\xi)\pi}{2}\]
with the positions\begin{eqnarray*}
 &  & \theta_{1}<\mu_{1}<\theta_{2}<\theta_{3}<\mu_{2}<\theta_{4}\\
\mbox{or} &  & \theta_{1}<\mu_{2}<\theta_{2}<\theta_{3}<\mu_{1}<\theta_{4}\\
\mbox{or} &  & \theta_{1}<\theta_{2}<\mu_{1},\mu_{2}<\theta_{3}<\theta_{4}\end{eqnarray*}
There are three $\mathcal{C}=+1$ eigenvectors. One has magnons of
the form\[
\lambda_{1}=\mu_{1}+\frac{i\pi}{2}\quad\lambda_{2}=\mu_{2}+\frac{i\pi}{2}\]
and another one with\[
\lambda_{1}=\mu_{1}+\frac{i(1+\xi)\pi}{2}\quad\lambda_{2}=\mu_{2}+\frac{i(1+\xi)\pi}{2}\]
with the typical positions\[
\theta_{1}<\mu_{1}<\theta_{2}<\theta_{3}<\mu_{2}<\theta_{4}\]
The last eigenvector is of the type\[
\lambda_{1}=\mu+\frac{i\pi(1+x)}{2}\qquad\lambda_{1}=\mu+\frac{i\pi(1-x)}{2}\]
where $\mu$ is exactly\[
\mu=\frac{\theta_{1}+\theta_{2}+\theta_{3}+\theta_{4}}{4}\]
and the fundamental range of $x$ is\[
0<x<\xi/2\]
given that the algebraic Bethe Ansatz has periodicity $i\pi\xi$ (and
the sign of $x$ does not matter).

\subsection{The density of states}

The quantization conditions for the multi-soliton states can be written
as follows:\begin{eqnarray}
\mathrm{e}^{iML\sinh\theta_{j}}\Lambda\left(\theta_{j},\left\{ \lambda_{1},\dots,\lambda_{r}\right\} |\left\{ \theta_{1},\dots,\theta_{N}\right\} \right) & = & 1\qquad,\qquad j=1,\dots,N\nonumber \\
\prod_{k=1}^{N}a(\lambda_{j}-\theta_{k})\prod_{k\neq j}^{r}\frac{a(\lambda_{k}-\lambda_{j})}{a(\lambda_{j}-\lambda_{k})} & = & 1\qquad\quad,\qquad j=1,\dots,r\label{eq:nondiagby}\end{eqnarray}
where $N$ is the number of solitonic particles and $\mathcal{Q}=N-2r$
is their total topological charge. The magnonic configuration as detailed
in the previous subsection fixes the {}``isospin'' structure of
the state in the $2^{N}$ dimensional internal charge space. The two
equations must be solved simultaneously. In principle, the positions
of the magnons are fixed in terms of the rapidities $\theta_{1},\dots,\theta_{N}$
so the density of states with a given $N$, $r$ and fixed isospin
structure can be computed from the quantization relations of the solitons\[
Q_{j}(\theta_{1},\dots\theta_{N}|\lambda_{1},\dots,\lambda_{r})=ML\sinh\theta_{j}-i\log\Lambda\left(\theta_{j},\left\{ \lambda_{1},\dots,\lambda_{r}\right\} |\left\{ \theta_{1},\dots,\theta_{N}\right\} \right)=2\pi I_{j}\]
by taking the Jacobi determinant of the rapidity $\mapsto$ quantum
numbers mapping \cite{Pozsgay:2007kn}\[
\rho(\theta_{1},\dots,\theta_{N})=\det\left(\frac{\partial Q_{j}}{\partial\theta_{k}}\right)_{j,k=1,\dots,N}\]
where it is understood that we differentiate also the dependence of
the $\lambda_{s}$ with respect to the rapidities. However, there
is an easier way to obtain the density of states by extending the
set of BY equations by those of the magnons, considering also the
magnonic quantization relations\[
\mathcal{M}_{j}(\theta_{1},\dots\theta_{N}|\lambda_{1},\dots,\lambda_{r})=-i\sum_{k=1}^{N}\log\, a(\lambda_{j}-\theta_{k})-i\sum_{k\neq j}^{r}\log\,\frac{a(\lambda_{k}-\lambda_{j})}{a(\lambda_{j}-\lambda_{k})}=0\]
Then one has the following result\begin{equation}
\rho(\theta_{1},\dots,\theta_{N})=\left.\left(\nicefrac{{\displaystyle \det\frac{\partial(Q,\mathcal{M})}{\partial(\theta,\lambda)}}}{{\displaystyle \det\frac{\partial\mathcal{M}}{\partial\lambda}}}\right)\right|_{\lambda_{s}\rightarrow\lambda_{s}(\theta_{1},\dots,\theta_{N})}\label{eq:nondiagjacobi}\end{equation}
where the first determinant is the $(N+r)\times(N+r)$ Jacobi determinant
formed by including the magnonic equations, differentiating by keeping
the $\theta_{k}$ and $\lambda_{j}$ independent, and the second is
the $r\times r$ one containing only the magnonic relations. This
can be proven with a bit of labour by expressing the derivatives \[
\frac{\partial\lambda_{j}}{\partial\theta_{k}}\]
by differentiating the magnonic relations and using simple properties
of determinants. However, the result is very easy to understand intuitively.
Since the magnons do not have independently chosen quantum numbers
(their position is fixed by the rapidities of the solitons), the solitonic
phase space volume can be obtained from the phase space volume calculated
with all variables (rapidities and magnons) involved, divided by the
volume contribution of the magnons themselves.

\providecommand{\href}[2]{#2}\begingroup\raggedright
\endgroup
\end{document}